\documentclass[reqno]{amsart}
\usepackage{amssymb}
\usepackage[dvips]{epsfig}
\usepackage{graphicx}
\usepackage{color}
\usepackage{amsmath}
\usepackage{amssymb}
\usepackage{amsfonts}
\usepackage{amsthm}
\usepackage{mathrsfs}
\usepackage{graphicx,tikz}
\usepackage[colorlinks=true,pdfstartview=Fit,linkcolor=magenta,citecolor=cyan,urlcolor=cyan]{hyperref}
\definecolor{zeroblue}{RGB}{218,237,245}
\definecolor{tailblue}{RGB}{31,100,151}
\definecolor{cutred}{RGB}{171,57,47}
\definecolor{ink}{RGB}{53,65,74}
\definecolor{interiorbg}{RGB}{248,245,239}
\definecolor{tailbg}{RGB}{237,245,251}
\definecolor{caseblue}{RGB}{35,111,158}
\definecolor{caseorange}{RGB}{192,100,48}
\definecolor{zerocolor}{RGB}{188,36,115}

\newcommand{\R}{\mathbb R}

\newcommand{\C}{\mathbb C}

\newcommand{\T}{\mathbb{T}}

\newtheorem{thm}{Theorem}[section]
\newtheorem{lem}[thm]{Lemma}

\theoremstyle{remark}
\newtheorem{rem}{\bf Remark}[section]
\theoremstyle{definition}

\numberwithin{equation}{section}

\usepackage{bm}
\begin{document}

\title[ ]{Slow information spreading in quasi-periodic quantum spin chains}

	\author[Yingte Sun]{Yingte Sun*}
      \thanks{*Corrseponding author}

	\address{School of Mathematical Sciences, Yangzhou University, Yangzhou, P.R.China}
	\address{Spin Quantum Computing Laboratory, eSpinmag (Shanghai) Technology Co., Ltd, Shanghai, P.R.China} \email{sytsts@163.com}


\begin{abstract}
We investigate the suppression of information propagation in a weakly interacting quantum spin chain subject to a quasi-periodic external magnetic field. Using a finite-order normal-form construction, we establish a localized barrier in the interior of the chain that inhibits information propagation and derive a quantitative bound on the commutator of observables supported at opposite ends of the chain.
 \end{abstract}

\maketitle

\section{Introduction and main results}
\subsection{Motivation and relation to localization}\

Locality imposes fundamental constraints on the propagation of information in quantum lattice systems. For finite-range interactions, the Lieb–Robinson bound controls the commutator of separated local observables, thereby establishing an effective finite propagation speed \cite{LR1972}. However, additional structure in the system, such as a spatially inhomogeneous external field, may lead to stronger dynamical bounds. This possibility is closely related to the phenomenon of localization. In the single-particle setting, Anderson demonstrated that disorder can suppress transport \cite{A1958}. Similar localization phenomena can also arise in deterministic quasi-periodic potentials, where their occurrence depends on the coupling regime and the arithmetic properties of the frequency and phase. A prominent example is the almost Mathieu operator \cite{J99}.

For interacting quantum systems, the relationship between localization and information propagation becomes more subtle. A widely accepted description of the many-body localized regime involves a complete family of quasi-local integrals of motion (LIOMs) \cite{JVS17}. These are conserved operators $\widetilde{Z}_j$ concentrated near individual sites, with spatially decaying tails, satisfying $[H,\widetilde{Z}_j]=0$. Such conserved quantities provide a natural mechanism for suppressing the spreading of local information and motivate the study of improved Lieb–Robinson bounds. In this direction, Nachtergaele and Reschke \cite{NR2021} investigated the connection between local integrals of motion and dynamical localization in quantum lattice systems. Further examples are provided by uniform-in-time dynamical localization bounds for disordered XY chains \cite{HSS12} and sub-ballistic Lieb–Robinson bounds for certain quasi-periodic XY chains \cite{DLLY2014}. These results illustrate how localization phenomena can lead to slower information propagation. However, the results for XY chains  rely on integrable structures of the underlying models and can not directly extend to general nearest-neighbor interactions.

These considerations raise a fundamental question: \textbf{to what extent can a deterministic external field suppress information propagation in an interacting quantum many-body system without relying on any special structure of the interactions?} In this paper, we investigate this question in the setting of a weakly interacting quantum spin chain subject to a quasi-periodic external magnetic field, allowing arbitrary uniformly bounded nearest-neighbor interactions. By means of a finite-order normal-form transformation, we construct a localized barrier within the chain and show that,  outside a set of resonant frequencies of small measure, the end-to-end commutator bound acquires an additional suppression factor that is super power-law small in the interaction strength, compared with the conventional Lieb–Robinson bound. Although this result demonstrates the effectiveness of a deterministic external field in suppressing information propagation, the spatial scale of the barrier constructed here remains constrained by the smallness of the interaction strength. A further fundamental question is whether this restriction can be removed, so that effective barriers can be constructed on arbitrarily large spatial scales at a fixed nonzero interaction strength. Such a construction, together with suitable dynamical control, would open up the possibility of establishing a genuinely global suppression of information propagation, rather than merely inhibiting its transmission across a localized region. The present work constitutes a first step in this direction. 

\subsection{Model and dynamics}\

 Let $N \geq 3$ and consider the finite chain with open boundaries
$$
\Lambda_N:=\{1,2,\ldots,N\}, \quad  \quad 
\mathcal{H}_{\Lambda_N}:=\bigotimes_{k=1}^{N}\mathbb C^2.
$$
For $X \subseteq \Lambda_N$, set 
$\mathcal{H}_X := \bigotimes_{k\in X}\C^2, \  \mathcal{A}_X := \mathcal{B}(\mathcal{H}_X).
$
We identify an operator $A\in \mathcal{A}_X$ with $A\otimes \mathbb{I}_{\Lambda_N\setminus X}$ on the full Hilbert space. Throughout, \(\|\cdot\|\) denotes the operator norm.

The Hamiltonian is
\begin{equation}\label{ham}
	\mathrm{H}_{\Lambda_N}= \sum_{k=1}^N \cos(\theta+k\omega)\sigma_k^{(z)}+
	\epsilon \sum_{k=1}^{N-1} V_k,
\end{equation}
where $\theta\in \T:=\R/(2\pi\mathbb{Z})$ is the phase, $\omega\in[0,2\pi]$ is the frequency, and $\epsilon \in\R$ is the interaction strength. Here $\sigma_k^{(z)}$ is the Pauli matrix $\mathbf{diag}(1,-1)$ acting at site \(k\). Each $V_k\in  A_{\{k,k+1\}}$ is self-adjoint, and
\begin{equation}
	\sup_{1\le k\le N-1}\|V_k\|\le 1.
\end{equation}
When \(\omega/ 2\pi \) is irrational, the field is quasi-periodic in the spatial index; rational frequencies give periodic fields. The Hamiltonian is defined in both cases. 

We abbreviate  $\mathrm{H}_{\Lambda_N}$ to $\mathrm{H}$ and write
\begin{equation}
\mathrm{H}=H_0+\epsilon V,\qquad
H_0:=\sum_{k=1}^N \cos(\theta+k\omega)\sigma_k^{(z)},\qquad
V:=\sum_{k=1}^{N-1}V_k.
\end{equation}
Thus $H_0$ describes uncoupled spins, while $\epsilon V$ couples neighboring sites.

Since $\mathrm{H}$ is self-adjoint on the finite-dimensional  Hilbert space $\mathcal{H}_{\Lambda_N}$ . its Heisenberg dynamics  can be well defined by 
\begin{equation}
	\tau_t^{\mathrm{H}}(A):=e^{it\mathrm{H}}A e^{-it\mathrm{H}},\qquad t\in\R.
\end{equation}
For observables $A \in \mathcal{A}_X$ and $B\in \mathcal{A}_Y$ with disjoint supports, we study the commutator 
\begin{equation}\label{com}
	\left\|\left[\tau_t^\mathrm{H}(A),B\right]\right\|.
\end{equation}
This commutator vanishes at \(t=0\). Its norm controls the influence of an initially local observable on a distant one,  and serve us our  measure  of information spreading.

\subsection{Main result}\

The main result of this section is to demonstrate how, once a sufficiently long localized barrier can be constructed within the spin chain, it quantitative bounds for the commutator \eqref{com} as a function of the time $t$, the distance between the supports of $A$ and $B$, and the interaction strength $\epsilon$. This naturally brings into consideration a situation that cannot be overlooked: namely, when the spin chain length $N$ is not sufficiently large, so that the entire spin chain may itself be constructed as a single localized barrier. To address this case as well, we need to establish a simple preliminary lemma describing how the spin chain, after a unitary transformation, affects the Lieb-Robinson bounds.

\begin{lem}\label{lemma}\textbf{(Normal-form comparison} Let $N \geq 3$. Suppose $U$ is unitary and 
	\begin{equation}
		U\mathrm{H}U^*=\widetilde{H}_0+R, \quad  \widetilde{H}_0= \widetilde{H}_0^*, \quad R=R^*.
		\end{equation} 
	Then, for any local observables $A,B \in \mathcal{B}(\mathcal{H}_{\Lambda_N})$, take $\widetilde{A}=UAU^* $ and 
		$\widetilde{B}=UBU^* $. For any $t \in \mathbb{R}$,  one has 
			\begin{equation}
				\begin{split}
					\left\|[\tau^{\mathrm{H}}_{t}(A),B]\right\|  \leq \min \big\{ &2\|A\|\|B\|,\\
					&\|[\tau^{\widetilde{H}_0}_{t}(\widetilde{A}),\widetilde{B}]\|+4|t|\|R\|\|A\|\|B\| \big\}.\\
					\end{split}
				\end{equation}
	\end{lem}
\begin{proof}
	Since, 
	\begin{equation}
		\begin{split}
		Ue^{\mathrm{i}t\mathrm{H}}Ae^{-\mathrm{i}t\mathrm{H}}U^*&=U\left(A+[\mathrm{i}t\mathrm{H},A]+\sum^{\infty}_{n=2}\frac{\mathrm{ad}^n_{\mathrm{i}t\mathrm{H}}A}{n!}\right)U^*\\
		&=UAU^*+[\mathrm{i}t\widetilde{\mathrm{H}},\widetilde{A}]+\sum^{\infty}_{n=2}\frac{\mathrm{ad}^n_{\mathrm{i}t\widetilde{\mathrm{H}}}\widetilde{A}}{n!}\\
		&=e^{\mathrm{i}t\widetilde{\mathrm{H}}}Ae^{-\mathrm{i}t\widetilde{\mathrm{H}}},
	\end{split}
	\end{equation}
	then, one has 
	\begin{equation}
		U[\tau^{\mathrm{H}}_{t}(A),B]U^*=[\tau^{\widetilde{H}_0+R}_{t}(\widetilde{A}),\widetilde{B}].
		\end{equation}
		For any $t\geq0$, using Duhamel principle to obtain
		\begin{equation}
			\tau^{\widetilde{H}_0+R}_{t}(\widetilde{A})-\tau^{\widetilde{H}_0}_{t}(\widetilde{A})=\mathbf{i}\int^t_{0}\tau^{\widetilde{H}_0+R}_{s}([R, \tau^{\widetilde{H}_0}_{t-s}(\widetilde{A})])ds.
			\end{equation}
			Conequently,
			\begin{equation}
			\|\tau^{\widetilde{H}_0+R}_{t}(\widetilde{A})-\tau^{\widetilde{H}_0}_{t}(\widetilde{A})\| \leq 2|t|\|R\|\|A\|.
				\end{equation}
				Taking a commutator with $\widetilde{B}$ compelete the proof.
	\end{proof}

Accordingly, the main purpose of this paper is to understand how the quasi-periodic external field suppresses the spreading of local information generated by the interaction $V$, with particular emphasis on quantitative bounds for the commutator \eqref{com}
as a function of the time $t$, the distance between the supports of $A$ and $B$, and the interaction strength $\epsilon$.

\begin{thm}\label{main1}
Fix $0<\mu<\frac{1}{2}, 0<\beta<1$. There exists  $\epsilon^*=\epsilon^*(\beta,\mu)>0$ such tha the following assertions hold for $0 <\epsilon \leq \epsilon^*$,  each fixed phase $\theta \in \mathbb T$,  and
$$L=\ln(\frac{1}{\epsilon}), \quad \quad  m=\lfloor L^{\mu} \rfloor > 3.$$
Let $A \in \mathcal{A}_{\{1\}}$, $B  \in \mathcal{A}_{\{N\}}$, and $t>0$, then: 

\textbf{(I)Long chains}: If $N > 5m+1$,   there exisits a measurable resonant  set $\mathrm{R}_{\theta} \subseteq [0, 2\pi]$ satisfying

\begin{equation}
	\mathrm{meas} (\mathrm{R}_{\theta}) \leq  \exp \left[-\frac{1-\beta}{4} L^{1-\mu}\right].
\end{equation}
For $\omega \in [0,2\pi] \setminus \mathrm{R}_{\theta}$ ,  one has 
\begin{equation}\label{LRN}
	\left\|[\tau^{\mathrm{H}}_{t}(A),B]\right\|  \leq 20e \|A\| \|B\|  \epsilon^{\frac{1}{2}}\exp\left[-\frac{\beta}{2}L^{1+\mu}\right]t  \exp \left[20e\epsilon  t-r\right],
	\end{equation}
	where $$r=N-5m-1.$$
	
\textbf{(II) Short chians:} If $3< N \leq  5m+1$, there exisits a measurable resonant set $\mathrm{R}_{\theta} \subseteq [0, 2\pi]$ satisfying

\begin{equation}
	\mathrm{meas} (\mathrm{R}_{\theta}) \leq  
	\exp\left(-\frac{1-\beta}{4}L^{1-\mu}\right).
\end{equation}
For $\omega \in [0,2\pi] \setminus \mathrm{R}_{\theta}$ , 
\begin{equation}\label{SRN}
\|[\tau_t^H(A),B]\|
\leq \|A\|\|B\|
\min\left\{2,\,
4\epsilon |t|
\exp\left(-\frac{3\beta}{4}L
\left\lfloor\frac{N-2}{3}\right\rfloor\right)
\right\}.
\end{equation}
\end{thm}

\begin{rem}\
	
$\bullet$ For uniformly bounded finite-range interactions, the classical Lieb--Robinson bound \cite{LR1972,NS2019} takes the form
\begin{equation}\label{LR}
	\left\|[\tau_t^H(A),B]\right\|
	\le C_a\|A\|\|B\||X||Y|
	\exp\left[-a(d(X,Y)-v_a|t|)\right],
\end{equation}
where $a>0$, $d(X,Y):=\min\{|x-y|:x\in X,\ y\in Y\}$, and the constants are independent of $N$. In our nearest-neighbor model, passing to the interaction picture with respect to $H_0$ preserves the support and norm of each $V_k$, allowing the propagation velocity to be chosen as $v_a:=O(\epsilon)$. Thus, a velocity proportional to the interaction strength already follows from locality, while the improvement in \eqref{LRN} lies in its additional suppression factor.

$\bullet$ The restriction $\mu<\frac{1}{2}$ arises from the measure estimates for resonant frequencies, which require the decay in the measure bound to dominate the combinatorial growth of resonance conditions. This ensures that, for each fixed phase $\theta$, the non-resonance conditions hold outside a set of frequencies $\omega$ of small measure. Instead of the classical Wronskian method \cite{CSW25}, we employ the measure-theoretic Nazarov--Tur\'an inequality (Lemma \ref{app2}) for exponential polynomials. Its requirement of pairwise distinct frequencies is satisfied by the Fourier modes $\pm k$ appearing in our argument. 
\end{rem}

\subsection{Comment on the main result }\

	$\bullet$ A representative example covered by our framework is the XXZ quantum spin chain under  quasi-periodic magnetic field, a prototypical model for studying many-body localization:
	\begin{equation}
		H_{XXZ}
		=\epsilon_1\sum_{k=1}^{N-1}\sigma_k^{(z)}\sigma_{k+1}^{(z)}
		+\epsilon_2\sum_{k=1}^{N-1}
		\left(\sigma_k^{(x)}\sigma_{k+1}^{(x)}
		+\sigma_k^{(y)}\sigma_{k+1}^{(y)}\right)
		+\sum_{k=1}^{N}\cos(\theta+k\omega)\sigma_k^{(z)}.
	\end{equation}
	When the coupling constants $\epsilon_1$ and $\epsilon_2$ are sufficiently small, both the Ising and hopping interactions can be treated as perturbations of the uncoupled spin Hamiltonian. Thus, our results apply to this model in the weak-coupling regime, providing quantitative estimates on information propagation without requiring a proof of full many-body localization.
	
	$\bullet$ To obtain these estimates, we employ a finite-order normal-form transformation that eliminates non-diagonal perturbative terms within a suitable non-resonant region. An important feature of this approach is that the unperturbed on-site Hamiltonian remains fixed throughout the construction, allowing the non-resonance conditions to be formulated directly in terms of the original on-site energies. This makes the analysis of small divisors relatively transparent. The method is, however, limited by the combinatorial growth of higher-order terms, which prevents the present estimates from supporting an indefinite iteration. Related normal-form and block-diagonalization techniques have been employed by De Roeck et al. \cite{DHMP23} to investigate quasi-localization in disordered quantum chains, and by Fröhlich and Pizzo \cite{FP2020} to establish lower bounds on the spectral gap of quantum chains.
	
	$\bullet$ The difficulty of extending such constructions to arbitrarily large scales is closely related to the control of resonances, a central issue in the mathematical study of many-body localization. In this direction, De Roeck et al. \cite{DGHP25} recently developed a KAM-type iterative scheme and established the existence of a complete family of quasi-local integrals of motion for strongly disordered quantum chains on a non-resonant event whose probability is bounded below by an exponentially small quantity in the chain length. Their work illustrates how an all-scale analysis of resonances can lead to stronger localization conclusions. In the present quasi-periodic setting, the deterministic correlations of the external field pose a different challenge for controlling resonances across multiple scales. Rather than pursuing an infinite iteration, our approach relies on a finite-order construction of a localized barrier, which is sufficient to derive quantitative bounds on information propagation.

\section{Normal form transformation}

In this section, we outline a normal-form construction for quasi-periodic quantum spin chains. The transformed Hamiltonian exhibits an effective localized barrier to information propagation. The estimates needed to justify the construction and control the remainder, together with the measure estimates for the non-resonance conditions, are established in the following sections.

Let $m\geq 1$ and choose an integer $a$ such that $2\leq a$ and $a+5m\leq N$. Define the barrier and its subintervals by
\begin{equation}
\mathcal{S}:=\{a,\ldots,a+5m-1\},
\qquad
\mathcal{S}^{j}_{\ell}:=\{\ell ,\ldots,\ell+j\}.
\end{equation}
For fixed $\theta\in\mathbb{T}$ and $\delta>0$, define the non-resonant frequency set
\begin{equation}
	\mathrm{NR}_{\delta,m}(\theta;\mathcal{S})a
	:=
	\left\{
	\omega\in[0,2\pi]:
	\min_{\mathcal{S}^m_{\ell}\subseteq\mathcal{S}}  \min_{\mu\in\{-1,0,1\}^{\mathcal{S}^m_{\ell}}\setminus\{0\}
	}
	\left|
	2\sum_{k \in\mathcal{S}^m_{\ell}}
	\mu_\ell\cos(\theta+k \omega)
	\right|
	\geq\delta
	\right\}.
\end{equation}

\begin{thm}\label{th2.1}
	Suppose that $\omega\in\mathrm{NR}_{\delta,m}(\theta;\mathcal{S})$, and set
	\begin{equation}
	K_\delta=\frac{240}{\delta},
	\qquad
	\alpha=\frac{1}{60}.
	\end{equation}
	Then there exists a skew-adjoint operator
	\begin{equation}\label{skew}
		W=\sum_{j=1}^{m}\epsilon^j W^{(j)},
		\qquad
		\operatorname{supp}(W)\subseteq\mathcal{S},
	\end{equation}
	such that
	\begin{equation}\label{expa}
		\widetilde{\mathrm{H}}=e^W\mathrm{H}e^{-W}
		=H_0+\epsilon V_{\mathrm{out}}
		+\mathcal D+\mathcal F_L+\mathcal F_R+\mathcal R,
	\end{equation}
	where the terms satisfy the following properties:
	\begin{enumerate}
		\item The exterior interaction is
		\[
		V_{\mathrm{out}}
		:=
		\sum_{\substack{
				1\leq k\leq N-1\\
				\{k,k+1\}\cap\mathcal S=\varnothing
		}}V_k.
		\]
		
		\item  The  diagonal term $	\mathcal D$ has a  decomposition
		\begin{equation}\label{D-decomposition}
			\mathcal D
			=\sum_{j=1}^{m}\epsilon^j
			\sum_{\ell\in\mathcal I_j}\mathcal D_\ell^{(j)},
			\qquad
			\mathrm{supp}(\mathcal D_\ell^{(j)})
			\subseteq\mathcal S_\ell^{(j)},
		\end{equation}
		where $\mathcal I_j=\{a,\ldots, a+5m-1-j\}$, and $\mathcal D_\ell^{(j)}$ is self-adjoint and diagonal in
		the $\sigma^{(z)}$ basis.
		
		\item The boundary term $\mathcal F_L$ and $\mathcal F_R$ satisfy
		$$
		\mathrm{supp}(\mathcal F_L)
		\subseteq\{ a-1,\ldots, a+m\}, \quad 
		\mathrm{supp}(\mathcal F_R)
		\subseteq\{ a+4m-1,\ldots, a+5m\}.
		$$
		Moreover, if $|\epsilon|K_\delta m^3\leq\frac12$, then
		\begin{equation}
			\max\{\|\mathcal F_L\|,\|\mathcal F_R\|\}
			\leq 2 \epsilon.
		\end{equation}
		
		\item The higher-order remainder $\mathcal R$ is supported on
	$$
		\{a-1,\ldots, a+5m\}
	$$
		and admits the expansion
		\begin{equation}
			\mathcal R
			=\sum_{j=m+1}^{\infty}\epsilon^j\mathcal G^{(j)},
		\end{equation}
		where
		\begin{equation}
			\|\mathcal G^{(j)}\|
			\leq
			(6m+1)\frac{2}{e^{4\alpha}}
			\left(4K_\delta m^3e^{4\alpha}\right)^{j-1}.
		\end{equation}
	\end{enumerate}
\end{thm}
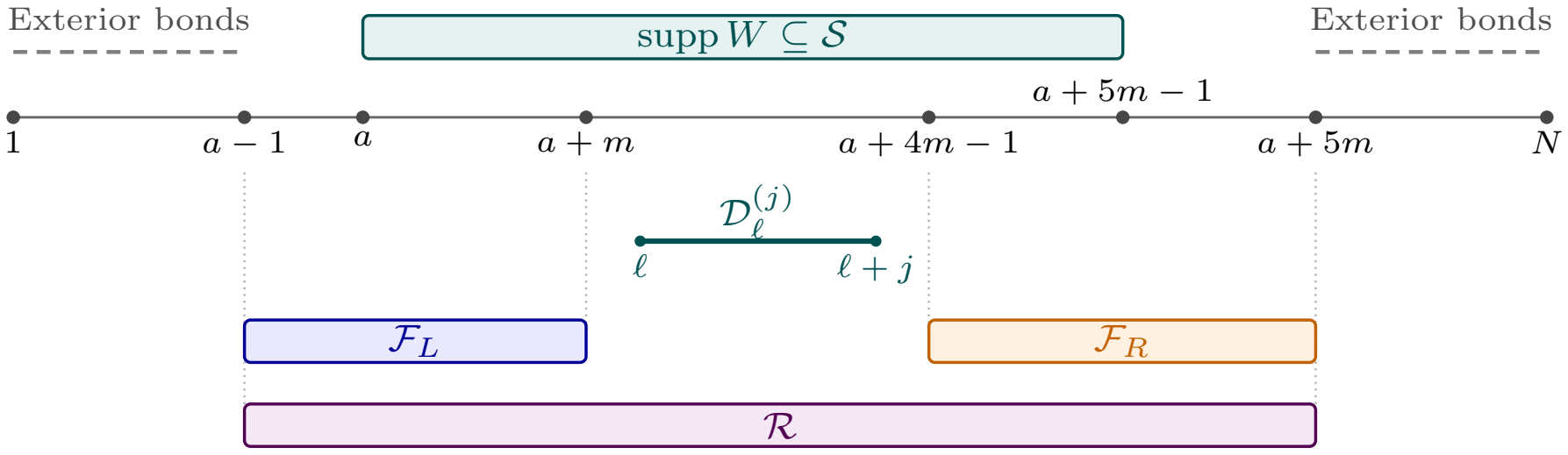
\begin{figure}[!htbp]
	\centering
	\begin{tikzpicture}[
		x=.85cm, y=.64cm, font=\footnotesize,
		site label/.style={font=\scriptsize, inner sep=1pt},
		guide/.style={black!30, densely dotted, line width=.45pt},
		outside/.style={draw=black!50, densely dashed, line width=.85pt},
		left support/.style={draw=blue!58!black, fill=blue!9},
		right support/.style={draw=orange!76!black, fill=orange!12},
		remainder/.style={draw=violet!65!black, fill=violet!9},
		support/.style={rounded corners=1.3pt, line width=.65pt}
		]
		\def\leftEdge{2.05}   
		\def\firstIn{3.10}    
		\def\leftEnd{5.08}    
		\def\rightStart{8.12} 
		\def\lastIn{9.84}     
		\def\rightEdge{11.55} 
		\def\chainEnd{13.60}  
		
		\draw[outside] (0,.77) -- (\leftEdge,.77);
		\draw[outside] (\rightEdge,.77) -- (\chainEnd,.77);
		\node[font=\scriptsize, text=black!70] at (1.025,1.16) {Exterior bonds};
		\node[font=\scriptsize, text=black!70] at (12.575,1.16) {Exterior bonds};
		\draw[support, draw=teal!64!black, fill=teal!10]
		(\firstIn,.69) rectangle (\lastIn,1.20);
		\node[text=teal!64!black] at (6.47,.945)
		{$\operatorname{supp} W\subseteq\mathcal S$};
		
		\draw[black!59, line width=.55pt] (0,0) -- (\chainEnd,0);
		\foreach \x/\name in {0/1,\leftEdge/{a-1},\firstIn/a,
			\leftEnd/{a+m},\rightStart/{a+4m-1},
			\rightEdge/{a+5m},\chainEnd/N} {
			\fill[black!72] (\x,0) circle[radius=1.45pt];
			\node[site label, anchor=north] at (\x,-.10) {$\name$};
		}
		\fill[black!72] (\lastIn,0) circle[radius=1.45pt];
		\node[site label, anchor=south] at (\lastIn,.08) {$a+5m-1$};
		
		\draw[teal!64!black, line width=1.15pt] (5.56,-1.46) -- (7.65,-1.46);
		\foreach \x/\name in {5.56/\ell,7.65/{\ell+j}} {
			\fill[teal!64!black] (\x,-1.46) circle[radius=1.2pt];
			\node[site label, text=teal!64!black, anchor=north]
			at (\x,-1.55) {$\name$};
		}
		\node[text=teal!64!black] at (6.605,-1.11)
		{$\mathcal D_\ell^{(j)}$};
		
		\foreach \x in {\leftEdge,\rightEdge}
		\draw[guide] (\x,-.65) -- (\x,-3.63);
		\foreach \x in {\leftEnd,\rightStart}
		\draw[guide] (\x,-.65) -- (\x,-2.64);
		
		\draw[support,left support]
		(\leftEdge,-2.89) rectangle (\leftEnd,-2.39);
		\node[text=blue!58!black] at (3.565,-2.64) {$\mathcal F_L$};
		\draw[support,right support]
		(\rightStart,-2.89) rectangle (\rightEdge,-2.39);
		\node[text=orange!76!black] at (9.835,-2.64) {$\mathcal F_R$};
		\draw[support,remainder]
		(\leftEdge,-3.88) rectangle (\rightEdge,-3.38);
		\node[text=violet!65!black] at (6.80,-3.63) {$\mathcal R$};
	\end{tikzpicture}
	\caption{Support bounds for the normal form. Here 
		$\mathcal S=\{a,\ldots, a+5m-1\}$. The segment below the chain shows a
		representative diagonal support bound $S_\ell^{(j)}=\{ \ell,\ldots, \ell+j\}$.
		Each term of $H_0$ acts on one site. Each exterior bond $V_k$
		in $V_{\mathrm{out}}$ acts on $\{k,k+1\}$ disjoint from $\mathcal S$.
		Site spacing is schematic.}
	\label{fig:normal-form-supports}
\end{figure}

\begin{proof}
	We first describe the recursive construction of $W$ at a formal level. The estimates required to justify the construction are established in Sections 3 and 4.
	
	Set
	\begin{equation}
	W=\sum_{j=1}^{m}\epsilon^jW^{(j)},
	\end{equation}
	and let $\operatorname{ad}_X(Y):=[X,Y]$. Then, the expansion \eqref{expa} gives
	\begin{equation}\label{w}
	\begin{split}
		\widetilde{\mathrm{H}}
		=&e^W(\mathrm{H}_0+\epsilon V)e^{-W}\\
	 =&e^{\sum_{j=1}^m\epsilon^iW^i}(\mathrm{H}_0+\epsilon V)e^{-\sum_{j=1}^m\epsilon^iW^i}\\ =&\mathrm{H}_0+\epsilon\Big([W^1,\mathrm{H}^0]+V \Big)\\ & \quad \ +\epsilon^2 \Big([W^2,H_0]+\frac{[W^1,[W^1,H^0]]}{2!}+[W^1,V] \Big)\\  &  \ \quad +\sum^{m}_{j=3}\epsilon^j \mathcal{G}^j + \mathcal{R}^{m+1}, \end{split} \end{equation}
		
	where, for $j\geq 2$
 \begin{equation}\label{G} 
 	\begin{split} \mathcal{G}^j=& \sum^{j-1}_{k=1}\frac{1}{k!}\sum_{\substack {r_1+\cdots r_k=j-1 \\ 1\leq r_1,\cdots r_k \leq \min(m,j-1)}}\mathrm{ad}_{W^{(r_k)}}\cdots \mathrm{ad}_{W^{(r_1)}}V\\ &\quad \quad +\quad \sum^{j}_{k=1}\frac{1}{k!}\sum_{\substack {r_1+\cdots r_k=j \\ 1\leq r_1,\cdots r_k \leq \min(m,j)}}\mathrm{ad}_{W^{(r_k)}}\cdots \mathrm{ad}_{W^{(r_1)}}H_0. 
 	\end{split} 
 \end{equation}
 
	For $1\leq j\leq m$, we separate the term containing $W^{(j)}$ and write
	\begin{equation}
		\mathcal G^{(j)}
		=[W^{(j)},H_0]+\mathcal F^{(j)},
	\end{equation}
	where $\mathcal F^{(j)}$ depends only on the generators of orders strictly below $j$.
	
	By locality, $\mathcal F^{(j)}$ can be decomposed as
	\begin{equation}
		\mathcal F^{(j)}
		=\mathbf{1}_{\{j=1\}}V
		+\mathcal F_L^{(j)}
		+\mathcal F_{\mathrm{In}}^{(j)}
		+\mathcal F_R^{(j)}.
	\end{equation}
	Here $\mathcal F_L^{(j)}$ and $\mathcal F_R^{(j)}$ collect the terms arising from the interactions across the left and right boundaries of $\mathcal S$, respectively. In particular,
	$$
	\mathcal F_L^{(1)}=V_{m^*-1},
	\qquad
	\mathcal F_R^{(1)}=V_{m^*+5m-1}.
	$$
	and for $j \geq 2$,
	\begin{equation}\label{bou1} F^j_{L}= \sum^{j-1}_{k=1}\frac{1}{k!}\sum_{\substack {r_1+\cdots r_k=j-1 \\ 1\leq r_1,\cdots r_k \leq \min(m,j-1)}}\mathrm{ad}_{W_{i_k}^{(r_k)}}\cdots \mathrm{ad}_{W_{i_1}^{(m_1)}}V_{m^*-1}, \end{equation} \begin{equation}\label{bou2} F^j_{R}= \sum^{j-1}_{k=1}\frac{1}{k!}\sum_{\substack {r_1+\cdots r_k=j-1 \\ 1\leq r_1,\cdots r_k \leq \min(m,j-1)}}\mathrm{ad}_{W_{i_k}^{(r_k)}}\cdots \mathrm{ad}_{W_{i_1}^{(r_1)}}V_{m^*+5m-1}, \end{equation} 
	The interior contribution can be expressed as
	\begin{equation}
		\mathcal F_{\mathrm{In}}^{(j)}
		=\sum_{\ell}\mathcal F_\ell^{(j)},
		\qquad
		\operatorname{supp}(\mathcal F_\ell^{(j)})
		\subseteq\mathcal S_\ell^{(j)},
	\end{equation}
	where $\mathcal{F}^1_\ell=V_{\ell}$, and for $j\geq 2$
\begin{equation}\label{eq1} 
	\begin{split}
		 F^j_{\ell}=&\quad \sum^{j-1}_{k=1}\frac{1}{k!}\sum_{\substack {r_1+\cdots r_k=j-1 \\ 1\leq r_1,\cdots r_k \leq \min(m,j-1) \\ \ell=\min(\ell_0,\cdots \ell_k)}}\mathrm{ad}_{W_{\ell_k}^{(m_k)}}\cdots \mathrm{ad}_{W_{\ell_1}^{(m_1)}}V_{\ell_0}\\ 
		 &\quad \quad +\quad \sum^{j}_{k=2}\frac{1}{k!}\sum_{\substack {r_1+\cdots r_k=j \\ 1\leq r_1,\cdots r_k \leq \min(m,j) \\ \min(\ell_0,\cdots \ell_k)=\ell}}\mathrm{ad}_{W_{\ell_k}^{(r_k)}}\cdots \mathrm{ad}_{W_{\ell_1}^{(r_1)}}(H_0)_{\ell_0}.
	  \end{split} \end{equation} 
	
	We now construct  the generator
$$
	W^{(j)}=\sum_\ell W_\ell^{(j)},
	$$
	to solve the homological equations
	\begin{equation}\label{hh}
		[W_\ell^{(j)},H_0]
		+\mathcal F_\ell^{(j)}
		-\mathcal D(\mathcal F_\ell^{(j)})=0,
		\end{equation}
such that
	\begin{equation}
		[W^{(j)},H_0]
		+\mathcal F_{\mathrm{In}}^{(j)}
		=\mathcal D(\mathcal F_{\mathrm{In}}^{(j)}).
	\end{equation}
	Collecting all terms up to order $m$, we obtain
	\begin{equation}
		\widetilde{\mathrm{H}}
		=H_0+\epsilon V_{\mathrm{out}}
		+\mathcal D+\mathcal F_L+\mathcal F_R+\mathcal R,
	\end{equation}
	where
	\begin{align}
		\mathcal D
		&=\sum_{j=1}^{m}
		\epsilon^j\mathcal D(\mathcal F_{\mathrm{In}}^{(j)})=\sum_{j=1}^{m}
		\epsilon^j \sum_{\ell \in \mathcal I_j}\mathcal D(\mathcal F_{\ell}^{(j)}),\\
		\mathcal F_L
		&=\sum_{j=1}^{m}\epsilon^j\mathcal F_L^{(j)},\quad  \quad
		\mathcal F_R
		=\sum_{j=1}^{m}\epsilon^j\mathcal F_R^{(j)},\\
		\mathcal R
		&=\sum_{j=m+1}^{\infty}
		\epsilon^j\mathcal G^{(j)}.
	\end{align}
	
	This completes the formal normal-form construction. The quantitative bounds on the generators, boundary terms, and remainder are established in Sections 3 and 4.
\end{proof}
\section{The homological equation and small divisors}
In the previous section, we outlined the normal-form transformation for the Hamiltonian $\mathrm{H}$. At each order $1\leq j\leq m$, our goal is to eliminate the off-diagonal part of $\mathcal F_{\mathrm{In}}^{(j)}$ while retaining its diagonal part. To this end, we consider the homological equation
\begin{equation}
	[W^{(j)},H_0]
	+\mathcal F_{\mathrm{In}}^{(j)}
	-\Pi_{\mathrm{diag}}
	\bigl(\mathcal F_{\mathrm{In}}^{(j)}\bigr)=0,
\end{equation}
where $\Pi_{\mathrm{diag}}$ denotes the projection onto operators diagonal in the $\sigma^{(z)}$ basis.

Recall that
$$
W^{(j)}
=\sum_{\ell\in\mathcal I_j}W_\ell^{(j)},
\qquad
\mathcal F_{\mathrm{In}}^{(j)}
=\sum_{\ell\in\mathcal I_j}\mathcal F_\ell^{(j)}.
$$
We therefore solve the local homological equations
\begin{equation}\label{he}
	[W_\ell^{(j)},H_0]
	+\mathcal F_\ell^{(j)}
	-\mathcal D_\ell^{(j)}=0,
\end{equation}
where
$$
\mathcal D_\ell^{(j)}
:=\Pi_{\mathrm{diag}}
\bigl(\mathcal F_\ell^{(j)}\bigr).
$$

To solve \eqref{he}, we work in the product eigenbasis of $H_0$. Let
$$
|\sigma\rangle
=\bigotimes_{k=1}^N|\sigma_k\rangle,
\qquad
\sigma_k\in\{-1,1\},
$$
where
$$
\sigma_k^{(z)}|\sigma\rangle
=\sigma_k|\sigma\rangle.
$$
Then
$$
H_0|\sigma\rangle
=E(\sigma)|\sigma\rangle,
\qquad
E(\sigma)
=\sum_{k=1}^N
\sigma_k\cos(\theta+k\omega).
$$

Since $\mathcal F_\ell^{(j)}$ is supported on
$\mathcal S_\ell^{(j)}$, its matrix elements vanish whenever two spin configurations differ outside this interval:
$$
\langle\sigma'|
\mathcal F_\ell^{(j)}
|\sigma\rangle=0,
$$
if
$$
\sigma_k'\neq\sigma_k,
\quad\text{for some }k\notin\mathcal S_\ell^{(j)}.
$$

Thus, only spin flips insider
$\mathcal S_\ell^{(j)}$ contribute to the homological equation. For such configurations,
\begin{equation}
	\begin{aligned}
		E(\sigma')-E(\sigma)
		&=\sum_{k\in\mathcal S_\ell^{(j)}}
		(\sigma_k'-\sigma_k)
		\cos(\theta+k\omega)\\
		&=2\sum_{k\in\mathcal S_\ell^{(j)}}
		\mu_k\cos(\theta+k\omega),
	\end{aligned}
\end{equation}
where
\[
\mu_k=\frac{\sigma_k'-\sigma_k}{2}
\in\{-1,0,1\}.
\]

This motivates the following local non-resonance condition:
\begin{equation}\label{eq:NR}
	\begin{aligned}
		\mathrm{NR}_{\delta,\mathcal S_\ell^{(j)}}(\theta)
		:=\Bigg\{
		\omega\in[0,2\pi]:\ 
		\min_{\substack{
				\mu\in\{-1,0,1\}^{\mathcal S_\ell^{(j)}\setminus\{0\}}
		}}
		\Bigg|
		2\sum_{k\in\mathcal S_\ell^{(j)}}
		\mu_k\cos(\theta+k\omega)
		\Bigg|
		\geq\delta
		\Bigg\}.
	\end{aligned}
\end{equation}

\begin{rem}
	The global non-resonance condition
	$\omega\in\mathrm{NR}_{\delta,m}(\theta;\mathcal S)$
	implies
	\[
	\omega\in
	\mathrm{NR}_{\delta,\mathcal S_\ell^{(j)}}(\theta)
	\]
	for every $1\leq j\leq m$ and
	$\ell\in\mathcal I_j$.
	Indeed, each $\mathcal S_\ell^{(j)}$ is contained in an interval of $m+1$ sites appearing in the definition of
	$\mathrm{NR}_{\delta,m}(\theta;\mathcal S)$, and every nonzero spin-flip pattern on
	$\mathcal S_\ell^{(j)}$ can be extended by zero to that interval.
\end{rem}

We now establish the solvability of the local homological equation and the corresponding norm estimates.

\begin{lem}\label{lem:homological}
	Assume that
	$\omega\in\mathrm{NR}_{\delta,m}(\theta;\mathcal S)$.
	For every $1\leq j\leq m$ and
	$\ell\in\mathcal I_j$, the homological equation
	\begin{equation}
		[W_\ell^{(j)},H_0]
		+\mathcal F_\ell^{(j)}
		-\mathcal D_\ell^{(j)}=0
	\end{equation}
	admits a unique solution supported on
	$\mathcal S_\ell^{(j)}$, subject to the normalization that its diagonal matrix elements vanish.
	
	More precisely, define $\mathcal B=\mathcal S_\ell^{(j)}$, one has 
	
	\begin{equation}
		\langle\sigma'|W_\ell^{(j)}|\sigma\rangle
		=
		\begin{cases}
			\dfrac{
				\langle\sigma'|\mathcal F_\ell^{(j)}|\sigma\rangle
			}{
				E(\sigma')-E(\sigma)
			},
			&\substack{
				\sigma'_{\mathcal B^c}=\sigma_{\mathcal B^c},\\
				\sigma'_{\mathcal B}\neq\sigma_{\mathcal B},
			}\\[8pt]
			0,&\text{otherwise}.
		\end{cases}
	\end{equation}
	
	Furthermore,
	\begin{equation}\label{h3}
		\|\mathcal D_\ell^{(j)}\|
			\leq\|\mathcal F_\ell^{(j)}\|,\quad
			\|\mathcal F_\ell^{(j)}-\mathcal D_\ell^{(j)}\|
			\leq2\|\mathcal F_\ell^{(j)}\|,
				\end{equation}
	\begin{equation}\label{h4}
		\|W_\ell^{(j)}\|
		\leq\frac4\delta
		\|\mathcal F_\ell^{(j)}\|.
	\end{equation}	If $\mathcal F_\ell^{(j)}$ is self-adjoint, then
	$W_\ell^{(j)}$ is skew-adjoint.
\end{lem}

\begin{proof}
	Set $\mathcal B=\mathcal S_\ell^{(j)}$.
	Since $\mathcal F_\ell^{(j)}$ is supported on
	$\mathcal B$, it suffices to consider the local Hilbert space
	$$
	\mathcal H_{\mathcal B}
	=\bigotimes_{k\in\mathcal B}\mathbb C^2.
	$$
	On this space, the unperturbed Hamiltonian is
	$$
	H_{0,\mathcal B}
	=\sum_{k\in\mathcal B}
	\cos(\theta+k\omega)\sigma_k^{(z)}.
	$$
	
	The non-resonance condition implies that any two distinct eigenvalues of
	$H_{0,\mathcal B}$ differ by at least $\delta$.
	Consequently, the off-diagonal matrix elements of \eqref{he} uniquely determine
	$W_\ell^{(j)}$, while its diagonal matrix elements are set to zero.

	The diagonal projection is contractive in the operator norm, so
	$$
	\|\mathcal D_\ell^{(j)}\|
	=\|\Pi_{\mathrm{diag}}
	(\mathcal F_\ell^{(j)})\|
	\leq\|\mathcal F_\ell^{(j)}\|.
	$$
	The second estimate in \eqref{h3} follows from the triangle inequality.
	
	We next prove \eqref{h4}.
	Let $d=2^{|\mathcal B|}$, one can arrange the eigenvalues of
	$H_{0,\mathcal B}$ in increasing order:
	\[
	\lambda_1<\lambda_2<\cdots<\lambda_d.
	\]
	By non-resonance,
	\[
	|\lambda_a-\lambda_b|
	\geq |a-b|\delta,
	\qquad a\neq b.
	\]
	Hence, for every $a$,
	\begin{equation}\label{eq:divisor-sum}
		\begin{aligned}
			\sum_{b\neq a}
			\frac{1}{|\lambda_a-\lambda_b|^2}
			&\leq\frac{2}{\delta^2}
			\sum_{n=1}^{\infty}\frac1{n^2}\\
			&=\frac{\pi^2}{3\delta^2}.
		\end{aligned}
	\end{equation}
	
	For convenience, we write 
	$$
	\mathcal{F}_{ab}
	=\langle a|\mathcal F_\ell^{(j)}|b\rangle,
	\qquad
	W_{ab}
	=\langle a|W_\ell^{(j)}|b\rangle.
	$$
	Then
$$
	W_{ab}
	=\frac{\mathcal{F}_{ab}}{\lambda_a-\lambda_b}
	\quad(a\neq b),
	\qquad W_{aa}=0.
$$
	
	For any vector $x=(x_1,\ldots,x_d)\in\mathbb C^d$, the Cauchy-Schwarz inequality gives
	\begin{align*}
		\|W_\ell^{(j)}x\|^2
		&=\sum_{a=1}^d
		\left|
		\sum_{b\neq a}
		\frac{\mathcal{F}_{ab}x_b}{\lambda_a-\lambda_b}
		\right|^2\\
		&\leq
		\sum_{a=1}^d
		\left(
		\sum_{b\neq a}
		\frac1{|\lambda_a-\lambda_b|^2}
		\right)
		\left(
		\sum_{b\neq a}|\mathcal{F}_{ab}x_b|^2
		\right)\\
		&\leq
		\frac{\pi^2}{3\delta^2}
		\sum_{b=1}^d
		|x_b|^2\sum_{a=1}^d|\mathcal F_{ab}|^2\\
		&\leq
		\frac{\pi^2}{3\delta^2}
		\|\mathcal F_\ell^{(j)}\|^2\|x\|^2.
	\end{align*}
	Therefore,
	$$
	\|W_\ell^{(j)}\|
	\leq\frac{\pi}{\sqrt3\,\delta}
	\|\mathcal F_\ell^{(j)}\|
	\leq\frac4\delta
	\|\mathcal F_\ell^{(j)}\|.
	$$
	
	Finally, if $\mathcal F_\ell^{(j)}$ is self-adjoint, then
	$$
	\overline{W_{ba}}
	=\frac{\overline{F_{ba}}}{\lambda_b-\lambda_a}
	=-\frac{F_{ab}}{\lambda_a-\lambda_b}
	=-W_{ab}.
	$$
	Thus
	$$
	(W_\ell^{(j)})^*=-W_\ell^{(j)}.
	$$
\end{proof}
\section{bounds on the reminder term}

In this section, we establish quantitative bounds on the normal-form generators $W^{(j)}$, the boundary terms $\mathcal F_L$ and $\mathcal F_R$, and the higher-order remainder $\mathcal R$. These estimates complete the quantitative part of Theorem \ref{th2.1}. The choice of non-resonace parameters and the resulting smallness of the remainder will be addressed in the next section.

\begin{lem}\label{l1}
	Assume that the non-resonance condition
	$\omega\in\mathrm{NR}_{\delta,m}(\theta;\mathcal S)$
	holds. Then, for each $1\leq j\leq m$, these constructed  skew-adjoint generators
	$$
	W^{(j)}
	=\sum_{\ell\in\mathcal I_j}W_\ell^{(j)}
	$$
 satisfy
	\begin{equation}\label{b1}
		\|\mathcal F_\ell^{(j)}\|
		\leq (j!)^3K_\delta^{j-1},
	\end{equation}
	and
	\begin{equation}\label{b2}
		\|W_\ell^{(j)}\|
		\leq\alpha(j!)^3K_\delta^j.
	\end{equation}

\end{lem}

\begin{proof}
	We proceed by induction on $j$.
	
	For $j=1$, the bounds follow from
$$
	\|\mathcal F_\ell^{(1)}\|
	=\|V_\ell\|\leq1
$$
	and Lemma \ref{lem:homological}, which gives
$$
	\|W_\ell^{(1)}\|
	\leq\frac4\delta
	=\alpha K_\delta.
$$
	
	Suppose that these estimates \eqref{b1} and \eqref{b2} hold for all orders $1\leq r<j$. By the local homological equation,
$$
	[W_\ell^{(r)},H_0]
	=\mathcal D_\ell^{(r)}
	-\mathcal F_\ell^{(r)}.
$$
	One can obtain
	\begin{equation}\label{eq:commutator-H0}
		\|[W_\ell^{(r)},H_0]\|
		\leq2\|\mathcal F_\ell^{(r)}\|.
	\end{equation}
	
	Using the expansion \eqref{eq1}, the induction hypothesis, we  can obtain
	\begin{align*}
		\|\mathcal F_\ell^{(j)}\|
		\leq{}&
		K_\delta^{j-1}
		\sum_{k=1}^{j-1}\frac{2^k\alpha^k}{k!}
		\sum_{\substack{
				r_1+\cdots+r_k=j-1\\
				r_i\geq1
		}}
		\prod_{i=1}^k(r_i!)^3
		N(1,r_1,\ldots,r_k)\\
		&+
		K_\delta^{j-1}
		\sum_{k=2}^{j}\frac{2^k\alpha^{k-1}}{k!}
		\sum_{\substack{
				r_1+\cdots+r_k=j\\
				r_i\geq1
		}}
		\prod_{i=1}^k(r_i!)^3
		N(r_1,\ldots,r_k).
	\end{align*}
	Here $N(\cdot)$ denotes the number of admissible intersecting interval chains.
	
	By the counting lemma \ref{countlem}, one has 
	\begin{align*}
		N(1,r_1,\ldots,r_k)
		&\leq
		2^{k+1}\frac{j\,j!}{(j-k)!},\\
		N(r_1,\ldots,r_k)
		&\leq
		2^k\frac{j\,j!}{(j-k+1)!}.
	\end{align*}
	Moreover, for positive integers $r_1,\ldots,r_k$,
	\begin{equation}\label{eq:factorial-bound}
		\prod_{i=1}^k r_i!
		\leq
		\left(\sum_{i=1}^k r_i-k+1\right)!.
	\end{equation}
	The number of compositions of an integer $j$ into $k$ positive parts is
$$
	\binom{j-1}{k-1}.
$$
	
	Combining these estimates yields
	\begin{align*}
		\|\mathcal F_\ell^{(j)}\|
		&\leq
		2(j!)^3K_\delta^{j-1}
		\sum_{k=1}^{j-1}
		\frac{(4\alpha)^k}{k!(k-1)!}
		+
		4(j!)^3K_\delta^{j-1}
		\sum_{k=1}^{j-1}
		\frac{(4\alpha)^k}{(k+1)!k!}\\
		&\leq
		(j!)^3K_\delta^{j-1}
		\left[
		2(e^{4\alpha}-1)
		+4(e^{4\alpha}-1)
		\right]\\
		&\leq(j!)^3K_\delta^{j-1},
		\end{align*}
	where the last inequality follows from
	$\alpha=1/60$.
	
	Finally, the estimate for the homological equation gives
$$
	\|W_\ell^{(j)}\|
	\leq\frac4\delta
	\|\mathcal F_\ell^{(j)}\|
	=\alpha K_\delta
	\|\mathcal F_\ell^{(j)}\|.
$$
	Using \eqref{b1}, we obtain
	$$
	\|W_\ell^{(j)}\|
	\leq\alpha(j!)^3K_\delta^j.
	$$
\end{proof}

\begin{lem}\label{l2}
	Let
	$$
	\mathcal F_L
	=\sum_{j=1}^m\epsilon^j\mathcal F_L^{(j)},
	\qquad
	\mathcal F_R
	=\sum_{j=1}^m\epsilon^j\mathcal F_R^{(j)}
	$$
	be the boundary terms defined in
	\eqref{bou1} and \eqref{bou2}.
	Then, for $2\leq j\leq m$,
	\begin{equation}\label{eq:boundary-coefficient}
		\max\left\{
		\epsilon^j\|\mathcal F_L^{(j)}\|,
		\epsilon^j\|\mathcal F_R^{(j)}\|
		\right\}
		\leq
		8\alpha\epsilon
		\left(\epsilon K_\delta j^3\right)^{j-1}.
	\end{equation}
\end{lem}

\begin{proof}
	We prove the estimate for $\mathcal F_L^{(j)}$; the proof for
	$\mathcal F_R^{(j)}$ is identical.
	
	For $j\geq2$, the boundary expansion  \eqref{bou1} and Lemma \ref{l1} give
	\begin{align*}
		\|\mathcal F_L^{(j)}\|
		\leq{}&
		K_\delta^{j-1}
		\sum_{k=1}^{j-1}\frac{(2\alpha)^k}{k!}
		\sum_{\substack{
				r_1+\cdots+r_k=j-1\\
				r_i\geq1
		}}
		N_{\partial}(r_1,\ldots,r_k)
		\prod_{i=1}^k(r_i!)^3.
	\end{align*}
	
	The boundary counting estimate gives
	$$
	N_{\partial}(r_1,\ldots,r_k)
	\leq2^k\frac{j!}{(j-k)!}.
$$
	Furthermore, since
	$r_1+\cdots+r_k=j-1$,
$$
	\prod_{i=1}^k(r_i!)^3
	\leq(j-1)^{3(j-k-1)}.
$$
	
	Hence,
	\begin{align*}
		\|\mathcal F_L^{(j)}\|
		\leq{}&
		K_\delta^{j-1}
		\sum_{k=1}^{j-1}
		\frac{(4\alpha)^k}{k!}
		\binom{j-2}{k-1}
		\frac{j!}{(j-k)!}
		(j-1)^{3(j-k-1)}.
	\end{align*}
	
	Using
	\[
	\binom{j-2}{k-1}
	\leq\frac{j^{k-1}}{(k-1)!},
	\qquad
	\frac{j!}{(j-k)!}\leq j^k,
	\]
	we obtain
	\begin{align*}
		\|\mathcal F_L^{(j)}\|
		&\leq
		K_\delta^{j-1}(j^3)^{j-1}
		\sum_{k=1}^{j-1}
		\frac{(4\alpha)^k}{k!(k-1)!}\\
		&\leq
		K_\delta^{j-1}(j^3)^{j-1}
		(e^{4\alpha}-1)\\
		&\leq
		8\alpha K_\delta^{j-1}(j^3)^{j-1}.
	\end{align*}
	Multiplying by $\epsilon^j$ yields
	\eqref{eq:boundary-coefficient}.
\end{proof}

\begin{rem}
	The first-order boundary terms satisfy
$$
	\mathcal F_L^{(1)}=V_{m^*-1},
	\qquad
	\mathcal F_R^{(1)}=V_{m^*+5m-1}.
$$
	If
$$
	\epsilon K_\delta m^3\leq\frac12,
$$
	then Lemma \ref{l2} implies
	\begin{align*}
		\max\{\|\mathcal F_L\|,\|\mathcal F_R\|\}
		&\leq
		|\epsilon|
		+8\alpha|\epsilon|
		\sum_{j=2}^m
		(|\epsilon|K_\delta m^3)^{j-1}\\
		&\leq
		(1+8\alpha)|\epsilon|\\
		&\leq2|\epsilon|.
	\end{align*}
\end{rem}

\begin{lem}\label{l3}
	Assume $m\geq3$. Let the higher-order remainder be
	\begin{equation}\label{R}
		\mathcal R
		=\sum_{j=m+1}^{\infty}
		\epsilon^j\mathcal G^{(j)},
		\qquad
		\mathcal G^{(j)}
		=\sum_\ell\mathcal G_\ell^{(j)}.
	\end{equation}
	Then, for every $j\geq m+1$, one has 
	\begin{equation}\label{eq:remainder-coefficient}
		\|\mathcal G_\ell^{(j)}\|
		\leq
		\frac2{e^{4\alpha}}
		\left(4K_\delta m^3e^{4\alpha}\right)^{j-1}.
	\end{equation}
\end{lem}

\begin{proof}
	We use the expansion \eqref{G} together with
	the bounds in Lemma \ref{l1}.
	
	Since every generator has order at most $m$,
	the integers $r_i$ appearing in the expansion
	satisfy $1\leq r_i\leq m$. Thus
$$
	r_i!\leq m^{r_i-1},
$$
	which gives
	\begin{equation}\label{eq:remainder-factorial}
		\prod_{i=1}^k(r_i!)^3
		\leq m^{3(\sum_{i=1}^k r_i-k)}.
	\end{equation}
	
	For the terms involving $V$, we have
	$\sum_i r_i=j-1$ and $k\geq1$.
	For the terms involving $H_0$, we have
	$\sum_i r_i=j$ and $k\geq2$.
	In both cases,
$$
	\prod_{i=1}^k(r_i!)^3
	\leq m^{3(j-2)}.
$$
The counting Lemma \ref{countlem} also implies
$$
	N(1,r_1,\ldots,r_k)
	\leq(2j)^{k+1},
	\qquad
	N(r_1,\ldots,r_k)
	\leq(2j)^k.
$$
	Moreover, the number of compositions of
	an integer $s$ into positive parts is at most
	$2^{s-1}$.
	
	Applying these estimates to the two families
	of commutators in \eqref{G}, one has
	\begin{align*}
		\|\mathcal G_\ell^{(j)}\|
		\leq&
		K_\delta^{j-1}m^{3(j-2)}
		\Bigg[
		2^{j-2}(2j)
		\sum_{k=1}^{j-1}
		\frac{(4\alpha j)^k}{k!}+
		\frac{2^{j-1}}{\alpha}
		\sum_{k=2}^{j}
		\frac{(4\alpha j)^k}{k!}
		\Bigg]\\
		&\leq
		2^{j-1}K_\delta^{j-1}m^{3(j-2)}
		\Big[
		j(e^{4\alpha j}-1)+
		\alpha^{-1}
		(e^{4\alpha j}-1-4\alpha j)
		\Big]\\
		&\leq
		2^{j-1}K_\delta^{j-1}m^{3(j-2)}
		(j+\alpha^{-1})e^{4\alpha j}.
	\end{align*}
	
	Since $\alpha=1/60$, $m\geq3$, and
	$j\geq m+1$, we have
$$
	\frac{
		e^{8\alpha}(j+\alpha^{-1})
	}{
		2^j m^3
	}
	\leq1.
	$$
	Therefore,
	\begin{align*}
		\|\mathcal G_\ell^{(j)}\|
		&\leq
		\frac2{e^{4\alpha}}
		\left(4K_\delta m^3e^{4\alpha}\right)^{j-1}.
	\end{align*}
\end{proof}
				
\section{The  convergence  and non-resonance estimates}

The purpose of this section is twofold. First, we show
that, with a suitable choice of the normal-form order $m$,
the higher-order remainder is super power-law small
in $\epsilon$. Second, for each fixed phase $\theta$,
we prove that all the required non-resonance conditions
are satisfied outside a  frequency set  of small measure.

Throughout this section, we set
\begin{equation}\label{eq:parameters}
	L:=\ln\frac{1}{\epsilon},
	\qquad
	m:=\lfloor L^\mu\rfloor,
	\qquad
	\delta:=\epsilon^{1-\beta}.
\end{equation}

\subsection{Smallness of the remainder}

We first establish quantitative bounds on the
generators and the higher-order remainder.

\begin{lem}\label{lem:smallness}
	Fix $0<\beta<1$ and $0<\mu<1$.
	There exists $\epsilon_*=\epsilon_*(\beta,\mu)>0$
	such that, for every $0<\epsilon<\epsilon_*$,
	the following estimates hold:
	\begin{enumerate}
		\item For the generators $W_\ell^{(j)}$ defined
		in \eqref{skew}, one has
		\begin{equation}\label{eq:W-small}
			\epsilon^j\|W_\ell^{(j)}\|
			\leq
			\exp\left(
			-\frac{\beta}{2}Lj
			\right),
			\qquad 1\leq j\leq m.
		\end{equation}
		
		\item For the higher-order remainder
		$\mathcal R$ defined in \eqref{R}, one has
		\begin{equation}\label{eq:R-small}
			\|\mathcal R\|
			\leq
			\epsilon^{1/2}
			\exp\left(
			-\frac{\beta}{2}L^{1+\mu}
			\right).
		\end{equation}
	\end{enumerate}
\end{lem}

\begin{proof}
	Recall that
$$
	K=\frac{240}{\delta},
	\qquad
	\alpha=\frac{1}{60}.
$$
	Thus,
$$
	\epsilon K=240\epsilon^\beta.
$$
	
	Since $m\leq L^\mu$, for any fixed constant $C>0$, one has 
	$$
	Cm^3\epsilon^\beta
	\leq CL^{3\mu}\epsilon^\beta
	\leq\epsilon^{\beta/2},
$$
	provided that $\epsilon$ is sufficiently small.
	
	From Lemma \ref{l1}, we obtain
	\begin{align*}
		\epsilon^j\|W_\ell^{(j)}\|
		&\leq
		\alpha(j!)^3(\epsilon K)^j\leq
		\alpha(\epsilon K m^3)^j\\
		&\leq
		\alpha(240m^3\epsilon^\beta)^j\\
		&\leq
		\exp\left(-\frac{\beta}{2}Lj\right).
	\end{align*}

	We next estimate the higher-order remainder. Set
$$
	q:=4K\epsilon m^3e^{4\alpha}
	=960e^{4\alpha}m^3\epsilon^\beta \leq\epsilon^{\beta/2}\leq\frac12,
$$
since $\epsilon_*$ small enough.
	From Lemma \ref{l3} , for every $j\geq m+1$, one has 
$$
	\epsilon^j\|G_\ell^{(j)}\|
	\leq
	\frac{2\epsilon}{e^{4\alpha}}q^{j-1}.
$$
	Since $\mathcal G^{(j)}$ contains at most
	$5m+1$ local terms, we have
	\begin{align*}
		\|\mathcal R\|
		&\leq
		\sum_{j=m+1}^{\infty}
		\epsilon^j\|\mathcal G^{(j)}\|\\
		&\leq
		\frac{2(5m+1)\epsilon}{e^{4\alpha}}
		\sum_{j=m+1}^{\infty}q^{j-1}\\
		&\leq 
		\frac{4(5m+1)\epsilon}{e^{4\alpha}}
		q^m\\
		&\leq
		C(m+1)\epsilon
		\exp\left(-\frac{\beta}{2}Lm\right),
	\end{align*}
	where $C>0$ is independent of $\epsilon$.
	
	Using $m=\lfloor L^\mu\rfloor\geq L^\mu-1$,
	we obtain
	\begin{align*}
		\|\mathcal R\|
		&\leq
		C(m+1)\epsilon
		\exp\left(
		-\frac{\beta}{2}L(L^\mu-1)
		\right)\\
		&=
		C(m+1)\epsilon^{1-\beta/2}
		\exp\left(
		-\frac{\beta}{2}L^{1+\mu}
		\right)\\
	&\leq
	\epsilon^{1/2}
	\exp\left(
	-\frac{\beta}{2}L^{1+\mu}
	\right).
	\end{align*}

\end{proof}

\subsection{Non-resonant frequencies}

We now verify the non-resonance conditions
required for the normal-form construction.
For each fixed phase $\theta$, we show that
these conditions hold after excluding a set
of frequencies of sufficiently small measure.

Define 
$$
\mathcal S_{k}=\{k,\ldots,k+m\}.
$$
For each nonzero vector
$$
\mu\in
\{-1,0,1\}^{\mathcal S_k}\setminus\{0\},
$$
one has 
\begin{equation}\label{eq:P-mu}
	P_\mu(\omega)
	:=
	\sum_{\ell\in\mathcal S_k}
	\mu_\ell\cos(\theta+\ell\omega).
\end{equation}
The corresponding resonant frequency set is
\begin{equation}\label{eq:E-mu}
	E_{\theta,k,\mu}
	:=
	\left\{
	\omega\in[0,2\pi]:
	|P_\mu(\omega)|<\frac{\delta}{2}
	\right\}.
\end{equation}

We define the local resonant set by
\begin{equation}\label{eq:R-local}
	\mathrm R_{\theta,k}
	:=
	\bigcup_{\substack{
			\mu\in\{-1,0,1\}^{\mathcal S_k}\\
			\mu\neq0
	}}
	E_{\theta,k,\mu}.
\end{equation}

\begin{lem}\label{lem:local-resonance}
	There exists a constant $C>0$, independent
	of $\epsilon$, $\theta$, and $k$, such that
	\begin{equation}\label{eq:local-measure}
		\operatorname{meas}(\mathrm R_{\theta,k})
		\leq
		C3^{m+1}
		\exp\left(
		-\frac{1-\beta}{2m+1}L
		\right).
	\end{equation}
\end{lem}

\begin{proof}
	Fix a nonzero vector
	$\mu\in\{-1,0,1\}^{\mathcal S_k}$.
	By expanding the cosine functions,
	we can write
	\begin{equation}
		\begin{aligned}
			P_\mu(\omega)
			=&
			\frac{e^{\mathrm i\theta}}{2}
			\sum_{\ell\in\mathcal S_k}
			\mu_\ell e^{\mathrm i\ell\omega}+
			\frac{e^{-\mathrm i\theta}}{2}
			\sum_{\ell\in\mathcal S_k}
			\mu_\ell e^{-\mathrm i\ell\omega}.
		\end{aligned}
	\end{equation}
	
	Since the site indices are positive,
	the nonzero Fourier frequencies
	$\pm\ell$ are pairwise distinct.
	Thus, $P_\mu$ is an exponential polynomial
	with at most $2(m+1)$ nonzero terms.
	
	Moreover, since the vector $\mu\neq0$, there always exists
	$\ell_0\in\mathcal S_k$ such that
	$|\mu_{\ell_0}|=1$. Its Fourier coefficient
	satisfies
	\[
	\left|
	\frac{1}{2\pi}
	\int_0^{2\pi}
	P_\mu(\omega)e^{-\mathrm i\ell_0\omega}
	\,d\omega
	\right|
	=\frac12.
	\]
	Consequently,
	\begin{equation}\label{eq:P-lower}
		\sup_{\omega\in[0,2\pi]}
		|P_\mu(\omega)|
		\geq\frac12.
	\end{equation}
	
	Let $n_\mu$ denote the number of nonzero
	Fourier terms in $P_\mu$.
	Then
$$
	2\leq n_\mu\leq2m+2.
$$
	
	Applying the Nazarov--Tur\'an inequality
	(Lemma \ref{app2}) to the set 
	$E_{\theta,k,\mu}$, we obtain
	\begin{equation}
		\begin{aligned}
			\sup_{\omega\in[0,2\pi]}
			|P_\mu(\omega)|
			\leq&
			\left(
			\frac{C}{\operatorname{meas}
				(E_{\theta,k,\mu})}
			\right)^{n_\mu-1} \cdot
			\sup_{\omega\in E_{\theta,k,\mu}}
			|P_\mu(\omega)|.
		\end{aligned}
	\end{equation}
	By \eqref{eq:E-mu} and \eqref{eq:P-lower}, one has 
	\[
	\frac12
	\leq
	\left(
	\frac{C}{\operatorname{meas}
		(E_{\theta,k,\mu})}
	\right)^{n_\mu-1}
	\frac{\delta}{2}.
	\]
	It follows that
	\begin{align}
		\operatorname{meas}(E_{\theta,k,\mu})
		&\leq C\delta^{1/(n_\mu-1)}
		\nonumber\\
		&\leq C\delta^{1/(2m+1)}.
	\end{align}
	
	Since there are at most $3^{m+1}-1$
	nonzero vectors $\mu$, the union bound gives
	\begin{align*}
		\operatorname{meas}(\mathrm R_{\theta,k})
		&\leq
		\sum_{\substack{
				\mu\in\{-1,0,1\}^{\mathcal S_k}\\
				\mu\neq0
		}}
		\operatorname{meas}(E_{\theta,k,\mu})\\
		&\leq
		C3^{m+1}\delta^{1/(2m+1)}\\
		&=
		C3^{m+1}
		\exp\left(
		-\frac{1-\beta}{2m+1}L
		\right).
	\end{align*}
\end{proof}

We now consider all subintervals
$\mathcal S_k$ contained in the barrier
$\mathcal S$ and define
\begin{equation}\label{eq:R-global}
	\mathrm R_\theta
	:=
	\bigcup_{\substack{
			k:\mathcal S_k\subseteq\mathcal S
	}}
	\mathrm R_{\theta,k}.
\end{equation}

By construction,
\[
[0,2\pi]\setminus\mathrm R_\theta
=
\mathrm{NR}_{\delta,m}(\theta;\mathcal S).
\]

\begin{lem}\label{lem:global-resonance}
	Fix $0<\beta<1$ and $0<\mu<\frac12$.
	For sufficiently small $\epsilon$, one has
	\begin{equation}\label{eq:global-measure}
		\operatorname{meas}(\mathrm R_\theta)
		\leq
		\exp\left(
		-\frac{1-\beta}{4}L^{1-\mu}
		\right).
	\end{equation}
\end{lem}

\begin{proof}
	Recall that the barrier has length $5m$,
	\[
	\mathcal S
	=
	\{m^*,\ldots,m^*+5m-1\}.
	\]
	Since each $\mathcal S_k$ contains $m+1$
	sites, there are exactly $4m$ admissible
	subintervals satisfying
	$\mathcal S_k\subseteq\mathcal S$.
	
	By Lemma \ref{lem:local-resonance}
	and the union bound,
	\begin{align*}
		\operatorname{meas}(\mathrm R_\theta)
		&\leq
		\sum_{\substack{
				k:\mathcal S_k\subseteq\mathcal S
		}}
		\operatorname{meas}(\mathrm R_{\theta,k})\\
		&\leq
		4mC3^{m+1}
		\exp\left(
		-\frac{1-\beta}{2m+1}L
		\right)\\
			&\leq
		\exp\left(
		-\frac{1-\beta}{4}L^{1-\mu}
		\right).
	\end{align*}
	
The last inequality holds, since 
$$
	\frac{L}{2m+1}
	=
	\left(\frac12+o(1)\right)L^{1-\mu}
	$$
and 
$$
	(m+1)\ln3+\ln(4Cm)
	=o(L^{1-\mu}),
$$
if $0<\mu<\frac{1}{2}$.
	
\end{proof}

\section{Derivation of the Lieb--Robinson bound}

In this section, we combine the normal-form
construction and the estimates established
in the preceding sections to prove the
long-chain assertion of Theorem \ref{main1}.

The proof consists of four steps. We first
use the normal-form transformation to reduce
the dynamics to the effective Hamiltonian.
We then pass to the interaction picture
with respect to its diagonal part and derive
an integral inequality for the commutator.
Finally, we estimate the resulting expansion
by counting chains of overlapping interactions.

\begin{proof}
	Throughout the proof, we assume $t\geq0$.  Recall that 
$$
	\mathcal S=\{a,\ldots,a+5m-1\},
$$
	and assume $m>3$ and $N>5m+1$.
	
	\medskip
	\noindent
	\textbf{Step 1: Reduction to the the effective Hamiltonian.}
	
	By Theorem \ref{th2.1}, there exists a
	skew-adjoint operator $W$ supported on
	$\mathcal S$ such that
	\begin{equation}\label{eq:conjugated-H}
		\widetilde {\mathrm{H}}=e^W\mathrm{H}e^{-W}
		=H_0+\epsilon V_{\mathrm{out}}
		+\mathcal D+\mathcal F_L+\mathcal F_R
		+\mathcal R.
	\end{equation}
	
	Let $A\in\mathcal A_{\{1\}}$ and
	$B\in\mathcal A_{\{N\}}$.
	Since the barrier is disjoint from the
	supports of $A$ and $B$,
	$$
	[W,A]=[W,B]=0.
$$
	Consequently,
$$
	e^WAe^{-W}=A,
	\qquad
	e^WBe^{-W}=B.
$$
	
	Using the unitary equivalence of
	$\mathrm H$ and $\widetilde{\mathrm{H}}$, we obtain
	\begin{align}
		\|[\tau_t^{\mathrm H}(A),B]\|
		&=
		\|e^W[\tau_t^{\mathrm H}(A),B]e^{-W}\|
		=
		\|[\tau_t^{\widetilde{\mathrm{H}}}(A),B]\|.
		\label{eq:unitary-comparison}
	\end{align}
	
	Thus, it suffices to estimate the
	commutator for the effective Hamiltonian $\widetilde{\mathrm{H}}$.
	
	\medskip
	\noindent
	\textbf{Step 2: The interaction picture.}
	
	Set
	\begin{equation}\label{eq:effective-splitting}
		\widetilde{\mathrm{H}}=\widetilde H_0+\mathcal P,
		\qquad
		\widetilde H_0:=H_0+\mathcal D,
	\end{equation}
	where
$$
	\mathcal P
	=
	\epsilon V_{\mathrm{out}}
	+\mathcal F_L+\mathcal F_R+\mathcal R.
$$
	and 
$$
	\mathcal D
	=
	\sum_{j=1}^m\epsilon^j
	\sum_\ell\mathcal D_\ell^{(j)}.
$$
	See that $\mathcal D_\ell^{(j)}$ is diagonal
	in the $\sigma^{(z)}$ basis and supported
	on an interval of at most $m+1$ sites. In particular, all these diagonal
	operators commute with one another
	and with $H_0$.
	
	For any operator $C$ supported on a
	set $X$, the commutativity of the
	diagonal terms implies
	\begin{equation}\label{eq:diagonal-support}
		\begin{aligned}
			&\operatorname{supp}
			\left(
			e^{it\widetilde H_0}
			Ce^{-it\widetilde H_0}
			\right) \subseteq
			X\cup
			\bigcup_{\substack{
					1\leq j\leq m,\ \\ \ell:
					\mathcal S_\ell^{(j)}\cap X
					\neq\varnothing
			}}
			\mathcal S_\ell^{(j)}.
		\end{aligned}
	\end{equation}
	Thus, conjugation by
	$e^{it\widetilde H_0}$ can enlarge the
	support by at most $m$ sites, independently
	of $t$.
	
	In the present construction, one has 
$$
	\operatorname{supp}(\mathcal F_L)
	\subseteq\{a-1,\ldots,a+m\},
\quad 
	\operatorname{supp}(\mathcal F_R)
	\subseteq\{a+4m-1,\dots, a+5m\}.
$$
	It follows that
	\begin{equation}\label{eq:dressed-boundaries}
		\begin{aligned}
			\operatorname{supp}(\mathcal F_L(t))
			&\subseteq\{a-1,\ldots, a+2m\},\\
			\operatorname{supp}(\mathcal F_R(t))
			&\subseteq\{a+3m-1,\ldots, a+5m\},
		\end{aligned}
	\end{equation}
	where
$$
	\mathcal F_{L,R}(t)
	=
	e^{it\widetilde H_0}
	\mathcal F_{L,R}
	e^{-it\widetilde H_0}.
$$
	Since $m>3$, the two intervals in
	\eqref{eq:dressed-boundaries} are disjoint.
	
	Moreover, the exterior interactions
	have supports disjoint from $\mathcal S$,
	while
$$
	\operatorname{supp}(\mathcal R)
	\subseteq\{a-1,\ldots,a+5m\}.
$$
	Therefore,
	\begin{align*}
		\operatorname{supp}(V_k(t))
		\subseteq\operatorname{supp}(V_k),\quad
		\operatorname{supp}(\mathcal R(t))
		\subseteq\{a-1,\ldots,a+5m\},
	\end{align*}
	and 
$$
	\|V_k(t)\|=\|V_k\|,
	\qquad
	\|\mathcal F_{L,R}(t)\|
	=\|\mathcal F_{L,R}\|,
	\qquad
	\|\mathcal R(t)\|=\|\mathcal R\|.
$$
	
	Define
	\begin{equation}\label{eq:interaction-U}
		U(t):=
		e^{it\widetilde H_0}
		e^{-it\widetilde H}.
	\end{equation}
	Then, one sees 
	\begin{equation}
		i\partial_tU(t)=\mathcal P(t)U(t),
		\qquad U(0)=I,
	\end{equation}
	where
	\begin{equation}
		\mathcal P(t)
		=
		e^{it\widetilde H_0}
		\mathcal Pe^{-it\widetilde H_0}.
	\end{equation}
	
	For any local observable $C$, define the flow
$$
	\Gamma_t(C):=U(t)^*CU(t).
$$
	Also, set
$$
	A_t:=
	e^{it\widetilde H_0}
	Ae^{-it\widetilde H_0}.
$$
	Since $A$ is supported at the left
	endpoint, and $\mathcal D$ is supported
	inside the barrier, one has 
$$
	A_t\in\mathcal A_{\{1\}},
	\qquad
	\|A_t\|=\|A\|.
$$
	
	A direct calculation shows
	\[
	\tau_t^{\widetilde H}(A)
	=\Gamma_t(A_t).
	\]
	Hence,
	\begin{equation}\label{eq:picture-comparison}
		\|[\tau_t^{\widetilde H}(A),B]\|
		=
		\|[\Gamma_t(A_t),B]\|.
	\end{equation}
	\begin{figure}[!htbp]
		\centering
		\begin{tikzpicture}[
			x=.85cm, y=.64cm, font=\footnotesize,
			site label/.style={font=\scriptsize, inner sep=1pt},
			guide/.style={black!30, densely dotted, line width=.45pt},
			outside/.style={draw=black!50, densely dashed, line width=.85pt},
			left edge/.style={draw=blue!58!black},
			right edge/.style={draw=orange!76!black},
			remainder/.style={draw=violet!65!black, fill=violet!9},
			support/.style={rounded corners=1.3pt, line width=.65pt}
			]
			\def\leftEdge{2.05}    
			\def\leftEnd{4.08}     
			\def\leftDressed{5.91} 
			\def\rightDressed{7.70}
			\def\rightStart{9.58}  
			\def\rightEdge{11.55}  
			\def\chainEnd{13.60}   
			
			\draw[outside] (0,.69) -- (\leftEdge,.69);
			\draw[outside] (\rightEdge,.69) -- (\chainEnd,.69);
			\node[font=\scriptsize, text=black!70] at (1.025,1.09) {Exterior bonds};
			\node[font=\scriptsize, text=black!70] at (12.575,1.09) {Exterior bonds};
			\node[text=black!75, anchor=south] at (0,.08) {$A_t$};
			\node[text=black!75, anchor=south] at (\chainEnd,.08) {$B$};
			\draw[black!59, line width=.55pt] (0,0) -- (\chainEnd,0);
			\foreach \x/\name in {0/1,\leftEdge/{a-1},\leftEnd/{a+m},
				\leftDressed/{a+2m},\rightDressed/{a+3m-1},
				\rightStart/{a+4m-1},\rightEdge/{a+5m},\chainEnd/N} {
				\fill[black!72] (\x,0) circle[radius=1.45pt];
				\node[site label, anchor=north] at (\x,-.10) {$\name$};
			}
			
			\foreach \x in {\leftEdge,\rightEdge}
			\draw[guide] (\x,-.65) -- (\x,-4.40);
			\foreach \x in {\leftEnd,\rightStart}
			\draw[guide] (\x,-.65) -- (\x,-2.65);
			\foreach \x in {\leftDressed,\rightDressed}
			\draw[guide] (\x,-.65) -- (\x,-3.36);
			
			\node[anchor=west, inner sep=0pt, text=black!70] at (0,-1.28) {Before};
			\draw[support,left edge,fill=blue!9]
			(\leftEdge,-1.53) rectangle (\leftEnd,-1.03);
			\node[text=blue!58!black] at (3.065,-1.28) {$\mathcal F_L$};
			\draw[support,right edge,fill=orange!12]
			(\rightStart,-1.53) rectangle (\rightEdge,-1.03);
			\node[text=orange!76!black] at (10.565,-1.28) {$\mathcal F_R$};
			
			\node[anchor=west, inner sep=0pt, text=black!70] at (0,-2.49) {After};
			\draw[support,left edge,fill=blue!9]
			(\leftEdge,-2.76) rectangle (\leftDressed,-2.22);
			\path[fill=blue!21]
			(\leftEnd,-2.71) rectangle (\leftDressed-.04,-2.27);
			\draw[blue!58!black, densely dashed, line width=.50pt]
			(\leftEnd,-2.76) -- (\leftEnd,-2.22);
			\node[text=blue!58!black] at (3.065,-2.49) {$\mathcal F_L(t)$};
			\draw[support,right edge,fill=orange!12]
			(\rightDressed,-2.76) rectangle (\rightEdge,-2.22);
			\path[fill=orange!24]
			(\rightDressed+.04,-2.71) rectangle (\rightStart,-2.27);
			\draw[orange!76!black, densely dashed, line width=.50pt]
			(\rightStart,-2.76) -- (\rightStart,-2.22);
			\node[text=orange!76!black] at (10.565,-2.49) {$\mathcal F_R(t)$};
			
			\draw[<->, black!65, line width=.50pt]
			(\leftDressed,-3.33) -- (\rightDressed,-3.33);
			\node[site label, text=black!75, anchor=north] at (6.805,-3.43)
			{$\operatorname{dist}=m-1$};
			
			\draw[support,remainder]
			(\leftEdge,-4.65) rectangle (\rightEdge,-4.15);
			\node[text=violet!65!black] at (6.80,-4.40) {$\mathcal R(t)$};
		\end{tikzpicture}
		\caption{Support bounds in the interaction picture. }
		\label{fig:interaction-picture-supports}
	\end{figure}
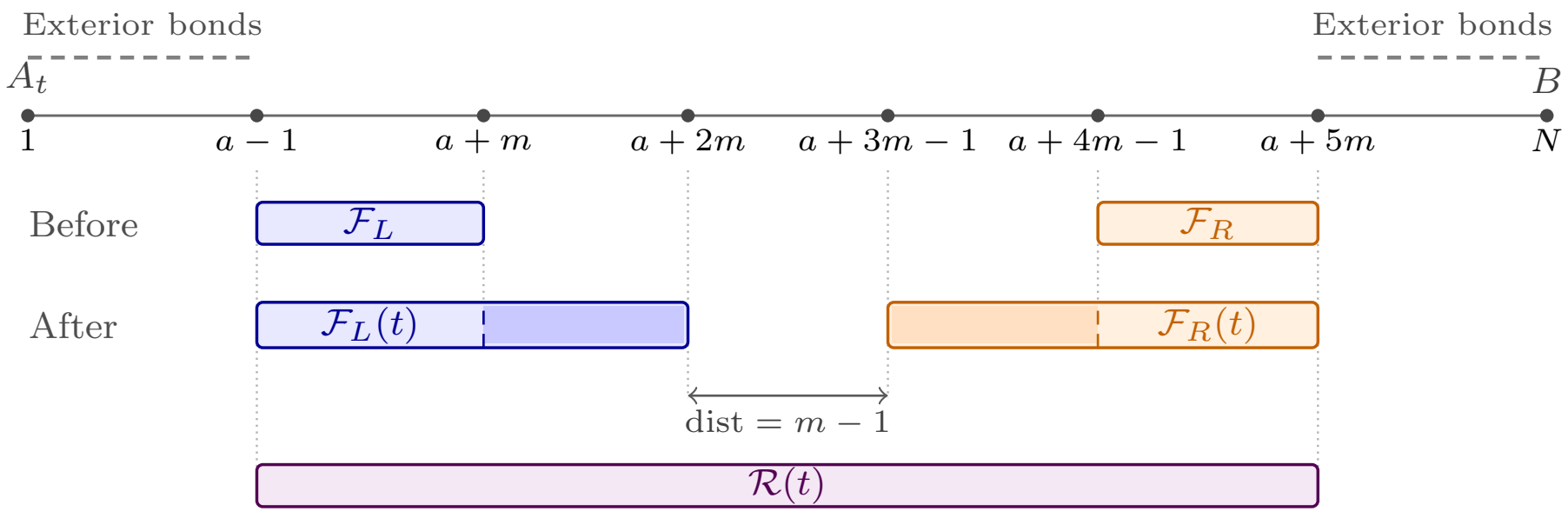
	\medskip
	\noindent
	\textbf{Step 3: An integral inequality.}
	The interaction-picture perturbation can be expressed as 
	\begin{equation}
		\mathcal P(t)= \sum_{e\in\mathscr E}\mathcal P_e(t),
	\end{equation}
	where the interaction labels $e$ correspond
	to the exterior bonds, the two boundary
	terms, and the  high order remainder.
	
	For each $e$, choose a fixed interval $X_e$
	containing the support of $\mathcal P_e(t)$
	for all $t$. The intervals may be chosen
	according to \eqref{eq:dressed-boundaries}
	and the support bounds established above.
	
	For a fixed local observable $B$, define
	\begin{equation}\label{eq:C-definition}
		C_B(X,s)
		:=
		\sup_{\substack{
				C\in\mathcal A_X\\
				C\neq0
		}}
		\frac{
			\|[\Gamma_s(C),B]\|
		}{
			\|C\|
		}.
	\end{equation}	Since
	$$
	\frac{d}{ds}\Gamma_s(C)
	=
	i\Gamma_s([\mathcal P(s),C]),
$$
	for any  $C\in\mathcal A_X$, one has 
	\begin{equation}\label{lie}
		\begin{aligned}
			\frac{d}{ds}[\Gamma_s(C),B]
			=&i\sum_{\substack{
					e\in\mathscr E:\\
					X_e\cap X\neq\varnothing
			}}
			[\Gamma_s([\mathcal P_e(s),C]),B]\\
			=&i\sum_{\substack{
					e\in\mathscr E:\\
					X_e\cap X\neq\varnothing
			}}
			[\Gamma_s(\mathcal P_e(s)),
			[\Gamma_s(C),B]]-
			i\sum_{\substack{
					e\in\mathscr E:\\
					X_e\cap X\neq\varnothing
			}}
			[\Gamma_s(C),
			[\Gamma_s(\mathcal P_e(s)),B]].
		\end{aligned}
	\end{equation}

	The first sum generates a unitary
	conjugation and therefore preserves
	the operator norm. Applying the
	Duhamel principle to
	\eqref{lie}, similar to Theorem A.1 in \cite{MKN2017}, we find
	\begin{align*}
		\|[\Gamma_t(C),B]\|
		\leq
		\|[C,B]\|+2\|C\|
		\sum_{\substack{
				e\in\mathscr E:\\
				X_e\cap X\neq\varnothing
		}}
		\int_0^t
		\|[\Gamma_s(\mathcal P_e(s)),B]\|
		\,ds.
	\end{align*}
	Since $\mathcal P_e(s)$ is supported
	on $X_e$ and
	$\|\mathcal P_e(s)\|=\|\mathcal P_e\|$,
	we obtain
	\begin{equation}\label{inin}
				C_B(X,t)
				\leq C_B(X,0)+2\sum_{\substack{
						e\in\mathscr E:\\
						X_e\cap X\neq\varnothing}}
				\|\mathcal P_e\|
				\int_0^t C_B(X_e,s)\,ds.
	\end{equation}
	
	\medskip
	\noindent
	\textbf{Step 4: Expansion and path counting.}
	
	Let
	\[
	X_A=\{1\},
	\qquad X_B=\{N\}.
	\]
	Since $X_A\cap X_B=\varnothing$,
	we have $C_B(X_A,0)=0$.
	
	Iterating \eqref{inin} gives
	\begin{equation}\label{eq:dyson-bound}
		C_B(X_A,t)
		\leq
		2\|B\|
		\sum_{n=1}^{\infty}
		\frac{(2t)^n}{n!}
		a_n(X_A,X_B),
	\end{equation}
	where
	\begin{equation}\label{eq:path-weight}
		\begin{aligned}
			a_n(X_A,X_B)
			:={}&
			\sum_{\substack{
					X_A\cap X_{e_1}\neq\varnothing\\
					X_{e_i}\cap X_{e_{i+1}}\neq\varnothing,
					\ 1\leq i<n\\
					X_{e_n}\cap X_B\neq\varnothing
			}}
			\prod_{i=1}^n\|\mathcal P_{e_i}\|.
		\end{aligned}
	\end{equation}
	
	The sum is taken over ordered chains
	of overlapping interaction supports
	connecting $X_A$ to $X_B$. We now use the geometry of the barrier.
	The dressed boundary supports satisfy
$$
	\{a-1,\ldots,a+2m\}
	\cap
	\{a+3m-1,\dots,a+5m\}
	=\varnothing.
$$
	Thus, without the remainder $\mathcal R$,
	the interaction supports on the left
	and right sides are disconnected.
	Consequently, every chain contributing
	to \eqref{eq:path-weight} must contain
	at least one occurrence of $\mathcal R$.
	
	Moreover, such a chain must traverse
	the exterior bonds between the left
	endpoint and the barrier, as well as
	those between the barrier and the
	right endpoint. Therefore its length satisfies
	\begin{equation}\label{eq:min-path-length}
		n\geq r,
		\qquad
		r:=N-5m-1.
	\end{equation}
	
	The overlap graph of the interaction
	supports has degree at most five when
	self-repetitions are included.
	Indeed, the remainder can overlap with
	the two boundary terms and the two
	adjacent exterior bonds, while all other
	interactions have fewer neighbors. Hence the number of admissible chains of length $n$ is bounded by $5^n$.
	
Thus, we obtain
	\begin{equation}\label{eq:path-estimate}
		a_n(X_A,X_B)
		\leq
		5^n(2\epsilon)^{n-1}
		\|\mathcal R\|,
		\qquad n\geq r.
	\end{equation}
	Substituting \eqref{eq:path-estimate}
	into \eqref{eq:dyson-bound}, we get
	\begin{align}
		C_B(X_A,t)
		&\leq
		2\|B\|\|\mathcal R\|
		\sum_{n=r}^{\infty}
		\frac{(2t)^n}{n!}
		5^n(2\epsilon)^{n-1}
		\nonumber\\
		&\leq 
		20t\|B\|\|\mathcal R\|
		\sum_{n=r}^{\infty}
		\frac{(20\epsilon t)^{n-1}}{n!}\nonumber\\
		&\leq
		20e\,t\|B\|\|\mathcal R\|
		\exp(20e\epsilon t-r).\nonumber\\
		\label{eq:intermediate-LR}
	\end{align}
	Since, for  $r\geq1$ and $x\geq0$,
	\begin{align*}
		\sum_{n=r}^{\infty}\frac{x^{n-1}}{n!}
		&\leq
		e^{-(r-1)}
		\sum_{n=r}^{\infty}
		\frac{(ex)^{n-1}}{n!}\\
		&\leq 
		e^{ex-(r-1)}.
	\end{align*}

\end{proof}

\section{Appendix}
\subsection{Remarks on short chains}\

The preceding section established the dynamical bound for long chains
by constructing a local normal form and comparing the resulting
dynamics. The same argument applies to short chains, with the interior
of the chain as the normal-form region and the normal-form order chosen according to its length.

	Suppose that $3< N\leq 5m+1$, and set
	\begin{equation}\label{short-chain-remark-parameters}
		\mathcal S_{\mathrm{sh}}:=\{2,\ldots,N-1\},
		\qquad
		p:=\left\lfloor\frac{N-2}{3}\right\rfloor.
	\end{equation}
	\smallskip
\textbf{Adaptation of the normal form:}
	Assume first that $p\geq1$. Impose the same non-resonance condition 
	$\mathrm{NR}_{\delta,p}(\theta;\mathcal S_{\mathrm{sh}})$ as before,
	but on every interval of $p+1$ sites contained in 
	$\mathcal S_{\mathrm{sh}}$.
	Every interval of at most $p+1$ sites in
	$\mathcal S_{\mathrm{sh}}$ is contained in one of
	these windows.
	
	The proof of Theorem~\ref{th2.1}, together with the
	estimates in Lemmas~\ref{l1}--\ref{l3}, applies with
	$m$ replaced by $p$ and $\mathcal S$ replaced by
	$\mathcal S_{\mathrm{sh}}$.  Restricting the allowed
	intervals to $\mathcal S_{\mathrm{sh}}$ does not increase
	the corresponding counting bounds. 
	
	Consequently, the same order-by-order construction gives
	a skew-adjoint generator
$$
	W_p=\sum_{j=1}^p\epsilon^j W^{(j)},
	\qquad
	\operatorname{supp}(W_p)\subseteq\mathcal S_{\mathrm{sh}},
$$
	such that
	\begin{equation}\label{short-chain-remark-nf}
		\widetilde{\mathrm{H}}_p:=e^{W_p}\mathrm{H}e^{-W_p}
		=H_0+\mathcal D+\mathcal F_L+\mathcal F_R+\mathcal R_p.
	\end{equation}
	There is no exterior interaction term, since every
	nearest-neighbor bond meets $\mathcal S_{\mathrm{sh}}$.
	The support estimates in the normal-form construction give
$$
	\operatorname{supp}(\mathcal F_L^{(j)})\subseteq\{1,\ldots,j+1\},
	\qquad
	\operatorname{supp}(\mathcal F_R^{(j)})\subseteq\{N-j,\ldots,N\},
	\qquad 1\leq j\leq p.
$$
	In particular, defining the interval
	$$
	X_L:=\{1,\ldots,p+1\}, \qquad X_R:=\{N-p,\ldots,N\},
	$$
	one has  $\operatorname{supp}(\mathcal F_L)\subseteq X_L$
	and $\operatorname{supp}(\mathcal F_R)\subseteq X_R$.
	Moreover,
$$
	\mathcal D
	=\sum_{j=1}^p\epsilon^j\sum_\ell\mathcal D_\ell^{(j)},
	\qquad
	\operatorname{supp}(\mathcal D_\ell^{(j)})
	\subseteq \{\ell,\ldots, \ell+j\},
$$
	where each $\mathcal D_\ell^{(j)}$ is self-adjoint
	and diagonal in the $\sigma^{(z)}$ basis.
	
	The
	local remainder bound of Lemma~\ref{l3} gives 
	\begin{equation}\label{short-chain-remark-remainder}
		\begin{aligned}
			\|\mathcal R_p\|
			\leq C N\epsilon\sum_{j=p+1}^{\infty} q_p^{\,j-1}, \quad \quad 
			q_p:
			=960e^{4\alpha}p^3\epsilon^\beta.
		\end{aligned}
	\end{equation}
Since
	$N\leq5L^\mu+1$ and $p\leq (5m-1)/3$, one sees  $q_p\leq1/2$
	for sufficiently small $\epsilon$. Furthermore,

	\begin{equation}\label{short-chain-remark-remainder-small}
		\|\mathcal R_p\|
		\leq 2C N(960e^{4\alpha}p^3)^p
		\leq \exp(-\frac{3}{4}\beta Lp).
	\end{equation}
	
	\smallskip
	\textbf{The geometric modification:}
	The only difference in the dynamical argument is the
	shorter barrier length. By \eqref{short-chain-remark-parameters},
$$
	\operatorname{dist}(X_L,X_R)=N-2p-1>p.
$$
	Hence the boundary supports are disjoint, and no individual
	diagonal term, whose support has diameter at most $p$,
	can meet both $X_L$ and $X_R$. 
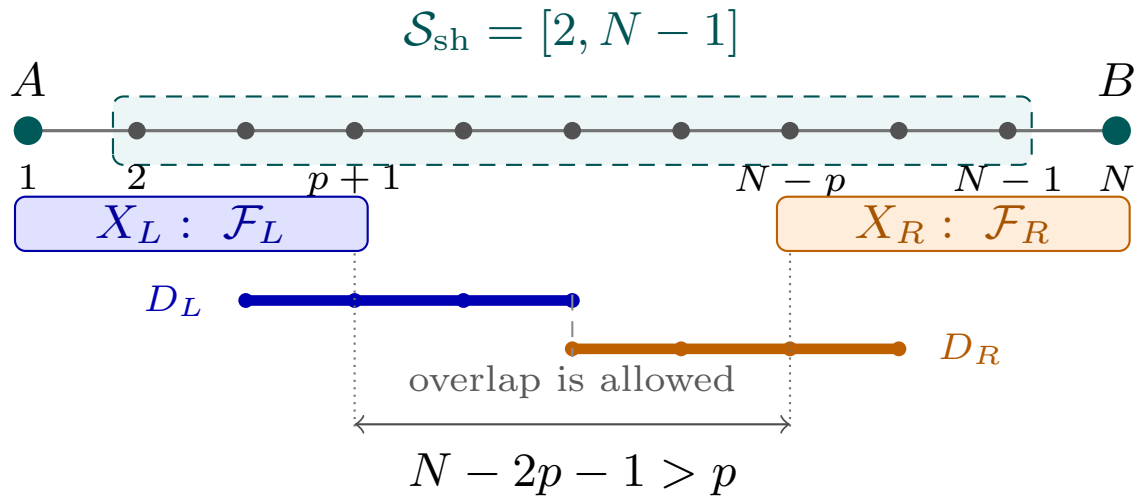
	\begin{center}
		\begin{tikzpicture}[x=.72cm,y=.78cm, font=\small]
			\fill[teal!7,rounded corners=2pt] (.78,-.29)
			rectangle (9.22,.29);
			\draw[teal!65!black,densely dashed,rounded corners=2pt]
			(.78,-.29) rectangle (9.22,.29);
			\draw[black!55,line width=.6pt] (0,0)--(10,0);
			\foreach \x in {0,...,10}{
				\fill[black!68] (\x,0) circle (1.65pt);
			}
			\fill[teal!70!black] (0,0) circle (2.7pt);
			\fill[teal!70!black] (10,0) circle (2.7pt);
			\node[above=3pt] at (0,0) {$A$};
			\node[above=3pt] at (10,0) {$B$};
			\node[above=4pt,text=teal!65!black] at (5,.29)
			{$\mathcal S_{\mathrm{sh}}=[2,N-1]$};
			
			\foreach \x/\lab in {0/1,1/2,3/{p+1},7/{N-p},
				9/{N-1},10/N}{
				\node[below=3pt,font=\scriptsize] at (\x,0) {$\lab$};
			}
			
			\fill[blue!12,rounded corners=2pt] (-.12,-1.02)
			rectangle (3.12,-.56);
			\draw[blue!65!black,rounded corners=2pt] (-.12,-1.02)
			rectangle (3.12,-.56);
			\node[text=blue!60!black] at (1.5,-.79)
			{$X_L:\ \mathcal F_L$};
			
			\fill[orange!14,rounded corners=2pt] (6.88,-1.02)
			rectangle (10.12,-.56);
			\draw[orange!75!black,rounded corners=2pt] (6.88,-1.02)
			rectangle (10.12,-.56);
			\node[text=orange!65!black] at (8.5,-.79)
			{$X_R:\ \mathcal F_R$};
			
			\draw[blue!70!black,line width=2.0pt,line cap=round]
			(2,-1.44)--(5,-1.44);
			\foreach \x in {2,3,4,5}
			\fill[blue!70!black] (\x,-1.44) circle (1.4pt);
			\node[anchor=east,font=\scriptsize,text=blue!70!black]
			at (1.81,-1.44) {$D_{L}$};
			
			\draw[orange!75!black,line width=2.0pt,line cap=round]
			(5,-1.85)--(8,-1.85);
			\foreach \x in {5,6,7,8}
			\fill[orange!75!black] (\x,-1.85) circle (1.4pt);
			\node[anchor=west,font=\scriptsize,text=orange!70!black]
			at (8.19,-1.85) {$D_{R}$};
			
			\draw[black!45,densely dashed] (5,-1.39)--(5,-1.92);
			\node[font=\scriptsize,text=black!65] at (5,-2.13)
			{overlap is allowed};
			
			\draw[black!55,thin,densely dotted] (3,-1.04)--(3,-2.52);
			\draw[black!55,thin,densely dotted] (7,-1.04)--(7,-2.52);
			\draw[<->,black!70] (3,-2.49)--(7,-2.49);
			\node[below=2pt] at (5,-2.49) {$N-2p-1>p$};
		\end{tikzpicture}
		
		\parbox{.9\linewidth}{\centering\footnotesize
			Illustration for $N=11$ and $p=3$. The two diagonal
			supports shown are representative terms of diameter at
			most $p$ that may overlap each other but cannot meet
			both boundary supports.}
	\end{center}
	
	Write $H_0+\mathcal D=D_L+D_M+D_R$, where $D_L$ and $D_R$
	collect the diagonal terms meeting $X_L$ and $X_R$,
	respectively, and $D_M$ collects the remaining terms
	(including the corresponding onsite terms of $H_0$).
	The three collections are well defined by the preceding
	geometric observation. Set
	\[
	H_L=D_L+\mathcal F_L,\qquad
	H_R=D_R+\mathcal F_R,\qquad
	H_M=D_M.
	\]
	By the commutativity of the diagonal terms and the
	support relations,
	\[
	[H_L,H_R]=[H_L,H_M]=[H_R,H_M]=0.
	\]
	In addition, $H_R$ and $H_M$ commute with $A$, while
	$H_L$ commutes with $B$. It follows that, for
	$H_{\mathrm{dec}}:=H_L+H_M+H_R$,
	\[
	\tau_t^{H_{\mathrm{dec}}}(A)=e^{itH_L}Ae^{-itH_L},
	\qquad
	[\tau_t^{H_{\mathrm{dec}}}(A),B]=0.
	\]
	This is the same decoupling mechanism used in the
	long-chain argument: the two ends cannot communicate
	after removal of the remainder.
	
	Since $W_p$ does not act on the two endpoints,
	$[W_p,A]=[W_p,B]=0$. Applying the Lemma \ref{lemma}  with
	$\widetilde H_p=H_{\mathrm{dec}}+\mathcal R_p$, gives

	\begin{align*}
		\|[\tau_t^H(A),B]\|&=\|[\tau_t^{\widetilde H_p}(A),B]\| \\
&\leq4|t|\|\mathcal R_p\|\|A\|\|B\|  \\
&\leq4\epsilon|t|e^{-\frac{3}{4}\beta L p}\|A\| \|B\| .
	\end{align*}
	
	Together with the trivial commutator bound, this proves
	\eqref{SRN}. 
	
	\smallskip
	\textbf{Resonant  frequencies.}
	The frequency estimate based on the Nazarov--Tur\'an
	inequality (Lemma~\ref{app2}) also holds with $p$ in place of $m$. For a fixed interval
	of $p+1$ sites, a nonzero coefficient vector in
	$\{-1,0,1\}^{p+1}$ defines an exponential polynomial with
	at most $2p+2$ distinct Fourier modes and supremum norm
	at least $1/2$. Taking the union over these vectors and
	the at most $N$ admissible intervals gives
	
	\begin{align*}
		\text{meas}(\mathrm R_{\theta,N})
		&\leq C N3^{p+1}\delta^{1/(2p+1)}\\
		&=C N\exp\left((p+1)\ln3
		-\frac{(1-\beta)L}{2p+1}\right)\\
		&\leq \exp \left(-\frac{1-\beta}{4}L^{1-\mu}\right),
	\end{align*}

Since $2p+1\leq(10m+1)/3$, and 
	$$
	\frac{L}{2p+1}
	\geq \left(\frac3{10}-o(1)\right)L^{1-\mu},
	\qquad
	(p+1)\ln3+\ln(C N)=o(L^{1-\mu}).
	$$

\subsection{Auxiliary lemmas}

\begin{lem}[Counting intersecting interval chains]
	\label{countlem}
	Let $I_1,\ldots,I_k$ be integer intervals,
	where $I_p$ contains $2r_p$ sites and
	$r_p\geq1$. Assume that
$$
	I_p\cap\bigcup_{q=1}^{p-1}I_q
	\neq\varnothing,
	\qquad 2\leq p\leq k.
$$
	If
$$
	r_1+\cdots+r_k=r
$$
	and the left endpoint of the final union
	is fixed, then the number of admissible
	ordered interval chains satisfies
	\begin{equation}\label{eq:chaincount}
		N(r_1,\ldots,r_k)
		\leq
		2^kr\frac{r!}{(r-k+1)!}.
	\end{equation}
	
	If instead the construction starts from
	a fixed boundary interval containing
	two sites, with
$$
	1+r_1+\cdots+r_k=r,
$$
	then
	\begin{equation}\label{eq:boundarychaincount}
		N_\partial(r_1,\ldots,r_k)
		\leq
		2^k\frac{r!}{(r-k)!}.
	\end{equation}
\end{lem}

\begin{proof} The proof follows that in \cite{DHMP23}.
	For completeness, we present the details below.
	
	Since each new interval intersects
	the union of the preceding intervals,
	the union remains an integer interval.
	For $p\geq2$, once the preceding intervals
	have been chosen, the left endpoint of
	$I_p$ has at most
$$
	2(r_1+\cdots+r_p)
$$
	admissible positions.
	
	When the left endpoint of the final
	union is fixed, the first interval
	has at most $2r$ possible translations.
	Consequently,
	$$
	N(r_1,\ldots,r_k)
	\leq
	(2r)\prod_{p=2}^k
	2(r_1+\cdots+r_p).
	$$
	
	The partial sums
$$
	r_1+r_2,\ldots,r_1+\cdots+r_k
$$
	are distinct positive integers not
	exceeding $r$. Therefore,
	$$
	\prod_{p=2}^k
	(r_1+\cdots+r_p)
	\leq
	\frac{r!}{(r-k+1)!}.
$$
	This proves \eqref{eq:chaincount}.
	
	For the boundary case, the initial
	interval is fixed. The left endpoint
	of the $p$th added interval has at most
$$
	2\left(1+\sum_{q=1}^p r_q\right)
$$
	admissible positions. Hence
$$
	N_\partial(r_1,\ldots,r_k)
	\leq
	2^k\prod_{p=1}^k
	\left(1+\sum_{q=1}^p r_q\right).
$$
	Since the $k$ partial sums are distinct
	positive integers not exceeding $r$,
$$
	\prod_{p=1}^k
	\left(1+\sum_{q=1}^p r_q\right)
	\leq
	\frac{r!}{(r-k)!}.
$$
	The second estimate follows.
\end{proof}

\begin{lem}[Nazarov--Tur\'an inequality, {\cite[Theorem 1.4]{Naz94}}]\label{app2}
	Let
	\[
	P(t)=\sum_{\nu=1}^{n}c_\nu e^{i\lambda_\nu t},
	\qquad c_\nu\in\mathbb{C}\setminus\{0\},
	\quad \lambda_\nu\in\mathbb{Z},
	\]
	where the frequencies $\lambda_\nu$ are pairwise distinct.
	Then there exists an absolute constant $C>0$ such that,
	for every measurable set $E\subseteq\mathbb{T}$
	of positive Lebesgue measure,
	\[
	\|P\|_{L^\infty(\mathbb{T})}
	\leq
	\left(\frac{C}{\operatorname{meas}(E)}\right)^{n-1}
	\|P\|_{L^\infty(E)}.
	\]

\end{lem}

\section*{Declarations}

	\textbf{AI statement} The first draft of this paper was completed in August 2026, and the author gave a  talk on this work at the end of August. No AI tools were used in writing the first draft. Later, the author used GPT-6.0 to generate two figures in this paper. GPT-6.0 also pointed out that the statement of the main theorem was not precise enough because the case of short chains was not considered. It also found an error in the use of the Nazarov--Tur\'an inequality. The author accepted these comments and made the corresponding corrections. The author takes full responsibility for all mathematical reesults in this paper.

\textbf{Funding} The author was supported by National Natural Science Foundation of China (Grant No.12101542).

\textbf{Conflict of interest}  On behalf of all authors, the corresponding author states that there is no conflict of interest.\\

\textbf{Availability of data }  The manuscript has no associated data.
\bibliographystyle{alpha} 
\bibliography{ref}

 \end{document}